\documentclass[fleqn,usenatbib]{mnras}

\usepackage{newtxtext,newtxmath}

\usepackage[T1]{fontenc}
\usepackage{ae,aecompl}

\usepackage{graphicx}	
\usepackage{amsmath}	
\usepackage{newtxtext,newtxmath}
\usepackage{amsfonts}

\title[Bursty H$\alpha$ Emitters]{Emission Line Galaxies in the SHARDS Frontier Fields I: Candidate Selection and the Discovery of Bursty H$\alpha$ Emitters}

\author[A. Griffiths~et~al.~]
{Alex~Griffiths,$^{1}$
Christopher~J.~Conselice$^{1,2}$\thanks{E-mail: \href{mailto:conselice@manchester.ac.uk}{conselice@manchester.ac.uk}},
Leonardo Ferreira$^{1}$,
Daniel Ceverino$^{3,4}$,
\newauthor
Daniel Rosa-Gonz\'{a}lez$^{5}$,
Marc Huertas-Company$^{6}$,
Bel\'{e}n Alcalde Pampliega$^{7,8}$,
\newauthor
Pablo G. P\'{e}rez-Gonz\'{a}lez$^{9}$,
Helena Dominguez Sanchez$^{10}$,
Olga Vega$^{5}$
\\
$^{1}$School of Physics and Astronomy, The University of Nottingham, University Park, Nottingham NG7 2RD, UK\\
$^{2}$Jodrell Bank Centre for Astrophysics, University of Manchester, Oxford Road, Manchester UK \\
$^{3}$Departamento de Fisica Teorica, Modulo 8, Facultad de Ciencias, Universidad Autonoma de Madrid, 28049 Madrid, Spain\\
$^{4}$CIAFF, Facultad de Ciencias, Universidad Autonoma de Madrid, 28049 Madrid, Spain\\
$^{5}$Instituto Nacional de Astrof\'{i}sica, \'{O}ptica y Electr\'{o}nica, AP 51 y 216, 72000, Puebla, M\'{e}xico \\
$^{6}$Departamento de Astrofisicia, Universidad de La Laguna, E-38206 La Laguna, Tenerife, Spain \\
$^{7}$European Southern Observatory (ESO), Alonso de C\'{o}rdova 3107, Vitacura, Casilla 19001, Santiago de Chile, Chile\\
$^{8}$Departamento de F\'{i}sica de la Tierra y Astrof\'{i}sica, Faultad de CC F\'{o}sicas, Universidad Complutense de Madrid E-2840 Madrid, Spain\\
$^{9}$Centro de Astrobiolog\'{\i}a (CAB, CSIC-INTA), Carretera deAjalvir km 4, E-28850 Torrej\'on de Ardoz, Madrid, Spain \\
$^{10}$Institute of Space Sciences (ICE, CSIC), Campus UAB, Carrer de Magrans, E-08193 Barcelona, Spain 
\\
}

\date{Accepted XXX. Received YYY; in original form ZZZ}

\pubyear{2021}

\begin{document}
\label{firstpage}
\pagerange{\pageref{firstpage}--\pageref{lastpage}}
\maketitle

\begin{abstract}

Emission line galaxies provide a crucial tool for the study of galaxy formation and evolution, providing a means to trace a galaxy's star formation history or metal enrichment, and to identify galaxies at a range of stellar masses.  In this paper we present a study of emission line galaxies in the SHARDS Frontier Fields medium-band survey. Through detailed flux calibrations we combine the first results of the SHARDS-FF survey with existing Hubble Frontier Field data to select 1,098 candidate emission line galaxies from the Hubble Frontier Filed clusters Abell 370 and MACS J1149.5+2223. Furthermore, we implement this deep medium-band imaging to update photometric redshift estimates and stellar population parameters and discover 38 predominantly low-mass H$\alpha$ emitters at redshifts $0.24 < z < 0.46$. Overall, 27 of these sources have corresponding UV data from the Hubble Space Telescope which allows us to distinguish these sources and investigate the burstiness of their star formation histories. We find that more than 50\% of our sample show an enhancement in H$\alpha$ over UV, suggesting recent bursts in star formation on time scales of a few, to tens of megayears. We investigate these sources and find that they are typically low-mass disky galaxies with normal sizes.  Their structures and star formation suggest that they are not undergoing mergers but are bursting due to alternative causes, such as gas accretion.

\end{abstract}





\section{Introduction}
\label{sec:intro}


Obtaining an accurate picture of galaxy formation and evolution are amongst the most outstanding questions in astrophysics today. There have been many studies examining this process in some detail over the past 30 years or so, including extensive Hubble Space Telescope campaigns \citep[e.g.][]{Conselice2003,Beckwith2006,Grogin2011,Duncan2019} and ground-based imaging studies \citep[e.g.][]{Steidel2004,Mundy2017}. One of the most critical aspects for studying galaxy formation and evolution is to determine the redshifts of objects such that their evolutionary connections can be made, either through mass or number density selections to connect systems at different points in their evolution \citep[e.g.][]{Mundy2015}.


Two major ways in which galaxy studies progress is through the investigation of galaxies at different redshifts, and by using the emission of light (direct or indirect) from these systems to measure the amount of current star formation and past assembly of stellar mass.  The process of star formation is fundamental for understanding the assembly of galaxies and how the universe was reionized. However, our resulting understanding of the star formation history of galaxies is largely based on observations of single integrated light measures of SFR within entire galaxies, over  a variety of redshifts \citep[e.g.][]{Madau2014}.   While much work in this area has been done, we are just starting to understand {\em how} and {\em why} star formation is distributed within galaxies.   We also know that the sources of reionization are likely galaxies at $z > 6$ \citep[e.g.][]{Duncan2015,Robertson2015}, yet we do not know from which types of galaxies, or modes of galaxy formation, ionizing radiation is emitted from.  Both of these problems can be addressed in unique ways by searching for and characterizing emission line systems in the distant universe, and perhaps by examining analogue galaxies at lower redshifts, such as the 'green peas' and Lyman-$\alpha$ emitters at $z \sim 1.5$ \citep[e.g.][]{Izotov2017,matthee2021}


There are in general three ways in which galaxies in the distant universe are discovered. The most reliable method is through spectroscopic surveys which, by accurately measuring the wavelengths of a few features in the spectrum of a galaxy, its redshift can be determined with a high accuracy \citep[e.g.][]{Zhou2019}.  However, spectroscopy of significant samples is still very difficult to obtain, and only future programs and telescopes such as MOONS \citep{moons2020}, WEAVE \citep{weave2020}, and 4MOST \citep{4most2019} will make significant progress towards obtaining significant numbers of redshifts for galaxies. Furthermore, the future of this field is 'blind' spectroscopy, with IFUs such as MUSE, where pre-selection is not required, as it is for most traditional spectroscopic surveys


Alternatively one can use photometric redshifts to locate the redshifts of distant galaxies by fitting intrinsic spectral energy distributions (SEDs) to stellar population models \citep[e.g.][]{Dahlen2013}. An alternative simplified version of this is to use the Lyman-break dropout method, whereby one finds the Lyman-break by searching for galaxies that have a faint, or statistically non-existing flux in an observed filter, traditionally the U-band \citep[e.g.][]{Koo1980,Guhathakurta1990,Steidel1993,Steidel2004}.

The final method is to use line emission as detected in narrow-band imaging. Searching for and examining distant galaxies using line emission through narrow-band imaging has a long history.  Traditionally, the Lyman-$\alpha$ line  is used to locate and study distant galaxies, \citep{rhoads2000,malhotra2004,Stark2010,Ouchi2010}. However, the radiative transfer of Ly-$\alpha$ is very complex \citep[e.g.][]{Gronke2016} as it is a resonant line, such that its photons scatter with neutral hydrogen, and increase the likelihood of dust absorption due to the short wavelengths. Thus, Ly-$\alpha$ cannot typically be used to trace the SFR in most galaxies, unless the escape fraction of Ly-$\alpha$  photons is known \citep[however, the work of][has calibrated Ly-$\alpha$ SFRs with rest frame equivalent widths to within 0.3 dex]{Sobral2019}.  It is often more direct and simplier to use the non-resonant H$\alpha$ \citep[e.g.][]{Geach2008,Sobral2013,Sobral2014}, or other similar lines such as [\ion{O}{II}] or even [\ion{O}{III}] \citep[e.g.][]{Khostovan2015}. Likewise, if two lines can be measured from the \ion{H}{ii} regions (such as \ion{O}{iii}/\ion{O}{ii}), their ratios gives an indication for the content of the radiation coming from these star forming areas \citep[e.g.][]{Izotov2017} and/or dust.


\subsection{H$\alpha$ Emitters}

For line emitters, Ly-$\alpha$ is ideal line for finding distant galaxies due to its brightness, but it is difficult to use this line for measuring astrophysical quantities from these galaxies \citep[see][]{Sobral2019}. The emission from H$\alpha$, [\ion{O}{III}] and [\ion{O}{II}] lines are, on the other hand, much less affected by dust than Ly-$\alpha$ or UV fluxes. These lines are also good for identifying and characterising galaxies which may have high dust attenuation and thus very faint Ly-$\alpha$ emission.   In this paper we present a general search for very faint line emitters using SHARDS \citep{Perez2013} data and describe a detailed analysis of the H$\alpha$ emitters we discover.  

There has been much progress in this over the past few years. For example, we know that bright H$\alpha$ emitters are very common at low and intermediate redshifts up to $z \sim 3$ \citep[e.g.][]{Sobral2012,Sobral2013}, but more difficult to find at higher-z due to technological limitations, as H$\alpha$ quickly redshifts into the NIR and redder wavelengths at modest redshifts.    We also have not yet probed the lowest mass galaxies at these epochs.  These may also be moderate redshift counter parts to the Lyman-continuum leaking Green Pea galaxies found in the local universe, for example \cite{Jaskot2013} and \cite{Yang2017}. They could also be analogues of the numerous  systems that reionized the universe \citep[e.g.,][Griffiths et al. 2021 submitted]{naidu2020}. 



Ground-based surveys have also used this H$\alpha$ line as a tracer of star formation in distant galaxies \citep[e.g.][]{Sobral2009,Sobral2012}, finding many sources and a steep luminosity function ($\alpha_{Ha} = -1.6$). H$\alpha$ can also be used to investigate the widely variable star formation rates (SFRs) of dwarf galaxies predicted by hydrodynamical simulations \citep[e.g.][]{Ceverino2016a,Ceverino2016b,Smit2016,Sparre2017,Emami2019}.  Dwarf galaxies are not just simple nearby systems, but are the most common galaxy type in the universe \citep[e.g.][]{Conselice2016}, but these are extremely difficult to locate and study in the distant Universe.  Deep emission line surveys remains one of the best ways to find and study these systems outside the very local Universe, with Ly-$\alpha$ surveys providing some of the lowest mass galaxies \citep{Oteo2015,Matthee2016,Santos2020}.



Dwarf galaxies with stellar masses of M$_* < 10^{9.5}$ M$_{\odot}$ typically experience bursty episodes of star formation on timescales of a few, to tens of megayears \citep{Rodighiero2011,Shen2014,Sparre2017}. In order to investigate the bursty star formation of these distant dwarf galaxies, it is necessary to use observables that trace star formation on different time scales. Hydrogen recombination lines such as H$\alpha$ and H$\beta$ are produced by short lived O-stars which have lifetimes of a few megayears such that their emission quickly reaches an equilibrium after the star formation is quenched, allowing us to trace galaxy SFRs on these short timescales. On the other hand, far-ultraviolet (FUV) continuum photons (1300\AA $< \lambda <$ 2000\AA) are produced by both O- and B-stars which have much longer lifetimes of ~ 100 Myr such that the FUV emission takes much longer to reach equilibrium. Thus, through the comparison of H$\alpha$ and FUV it is possible to determine variations in a galaxy's star formation rate on timescales of less than 100 Myr \citep[e.g.][]{Glazebrook1999,Iglesias2004,Lee2009,Weisz2012,Dominguez2015,Ceverino2018}.  This is crucial to determine how star formation occurs in these most common galaxies, which are also thought to be a dominate production of Lyman-continuum photons which reionized the universe \citep[e.g.,][Griffiths et al. 2021 submitted]{Duncan2015}. 

In this paper we describe a survey for these line emitters using the SHARDS Frontier Fields (FF) dataset obtained with the Gran Telescopio de Canarias (GTC).  The SHARDS-HFF survey is a medium band survey of the Frontier Fields, which we combine with the existing deep HST imaging.   Our data is unique in that our sources are found in this very deep medium-band imaging from in the Hubble-FF area.  We use this data to search for line emitters and to describe their basic properties and  reveal information about their star formation histories.   The purposes of this paper are two-fold: to describe our data and methodology for finding line emitters in the FFs, and to examine the nearest line emitters which are the H$\alpha$ systems.  Further papers in this series will describe the other line emitters, including an investigation of the Lyman-continuum emission for systems at $z \sim 2-3$ (Griffiths et al. 2021 submitted).


This paper is thus organised as follows: In Section~\ref{sec:obs} we discus the SHARDS medium-band observations and data reduction, as well as ancillary data used. Section~\ref{sec:calib} details the photometric calibration of SHARDS imaging matched to the deep HST broad-band data. Photometric redshifts and stellar population parameters are calculated in Section~\ref{sec:photz}. We describe corrections of the broad-band continuum colours and the selection of line emitters in Section~\ref{sec:cand}. We present result and investigate the properties of the H$\alpha$ emission line galaxies in Sections~\ref{sec:res} and \ref{sec:discussion}, and draw conclusions in Section~\ref{sec:conc}. Throughout this paper we adopt a $\Lambda$ cold dark matter cosmological model with $\Omega_{\Lambda} = 0.7$, $\Omega_{\textrm{M}} = 0.3$ and $H = 70$ km s$^{-1}$ Mpc$^{-1}$. All magnitudes are given in the AB system \citep{Oke1974}.


\section{Observations and data reduction}
\label{sec:obs}

The following analysis presented in this paper is based on new medium-band observations of the Hubble Frontier Fields (HFF) galaxy clusters \citep{Lotz2017}, Abell 370 and MACS J1149.5+2223 and their corresponding parallel fields. We also utilise the datasets, and multi-wavelength photometric catalogues made available through the HFF-DeepSpace project \citep{Shipley2018}. We explain the nature of this data and how we use it in our analyses in the following subsections.

\subsection{Data Sources}

\subsubsection{Medium-band observational data}

Medium-band (MB) optical imaging data of the Abell 370 and MACS J1149.5+2223 clusters (hereafter A0370 and M1149 respectively), as well as their parallel fields were obtained as part of the SHARDS Frontier Fields survey (SHARDS-FF; PI:P\'erez-Gonz\'alez). The SHARDS-FF (P\'erez-Gonz\'alez et al. in prep) survey is an ongoing observation program of two of the Hubble Frontier Fields (HFF) galaxy clusters, obtaining subarcsec-seeing imaging in 25 contiguous filters within the wavelength range 5000-9500~\AA, reaching an average spectral resolution $R\sim50$. Observations are performed using the Optical System for Imaging and low-Intermediate-Resolution Integrated Spectroscopy (OSIRIS) instrument, at the 10.4 m Gran Telescopio de Canarias (GTC) at the Observatorio del Roque de los Muchachos, in La Palma. OSIRIS's 8.5\arcmin x 7.8\arcmin~field of view (FOV) covers both the cluster and parallel HST fields in a single pointing. With observations beginning in December 2015.  A total of 240 hours observational time was granted for the program. On completion, the observations will reach at least 3-$\sigma$ sensitivity of m$\sim$27 in all the 25 contiguous medium-band filters. 

Individual images are reduced using a dedicated OSIRIS pipeline \citep{Perez2013}. The pipeline performs bias subtraction and flat fielding as well as illumination correction, background gradient subtraction and fringing removal. Additionally the pipeline implements World Coordinate System (WCS) alignment which includes field distortions, two-dimensional calibration of the passband and zero point and stacking of individual frames. 

\begin{figure*}
	\includegraphics{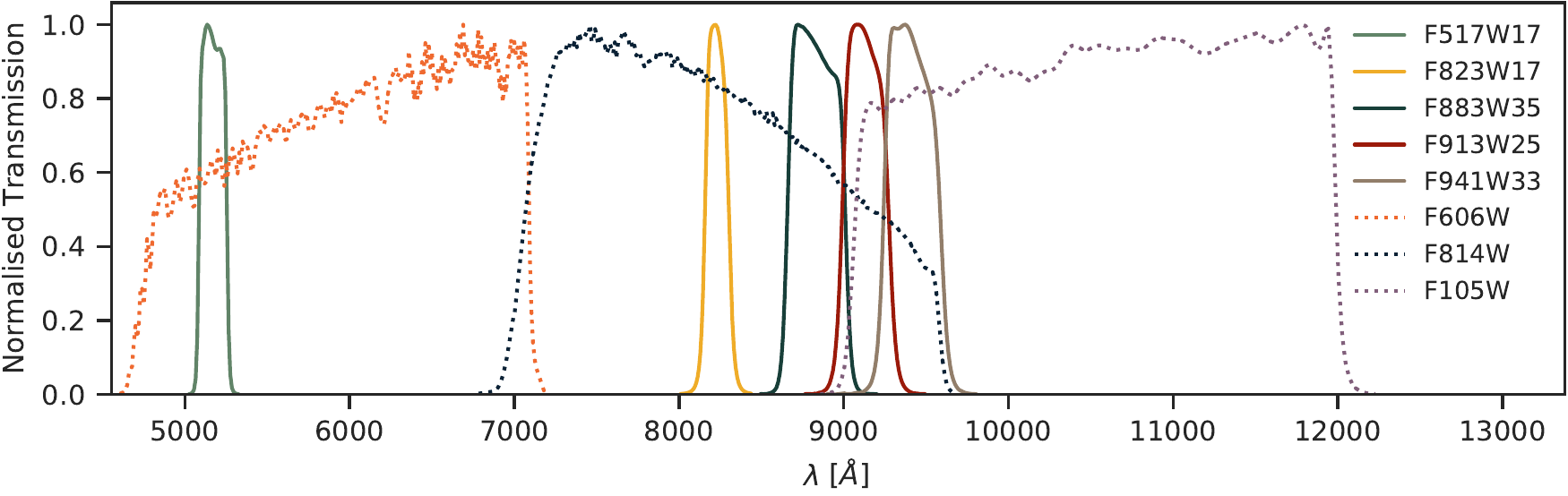}
	\caption{Transmission curves of the SHARDS and HST filters used for the selection of emission line galaxies. All curves have been individually normalised.}
	\label{fig:filters}
\end{figure*}

In order to search for emission line galaxies at a range of redshifts, we use all available SHARDS data; at the onset of this study, observations of A0370 have been carried out with four SHARDS filters (F517W17, F823W17, F913W25 and F941W33), while M1149 has been observed in three (F883W35, F913W25 and F941W33). Filter data is provided in Table~\ref{tab:filters} and we show response curves in Figure~\ref{fig:filters}. We take this opportunity to note that at the time of writing, SHARDS-FF observations of A0370 have recently been completed in all 25 contiguous medium-band filters in which raw data is publicly available as soon as the observations are completed. Reduced images and catalogues will be published in P{\'e}rez-Gonz{\'a}lez et al. (in prep.) and are available on direct request. 

\begin{table}
\caption{SHARDS Filter details.}
\label{tab:filters}
\centering
\begin{tabular}{ccc}
  \hline
  Filter  &  Central Wavelength  &  Width \\
          &       [\AA]          &  [\AA] \\
  \hline
  F517W17  &  5170  &  165  \\
  F823W17  &  8230  &  147  \\
  F883W35  &  8830  &  336  \\
  F913W25  &  9130  &  278  \\
  F941W33  &  9410  &  333  \\
  \hline
 \end{tabular}
\end{table}

\subsubsection{Ancillary observations and catalogues}

\begin{figure*}
	\includegraphics{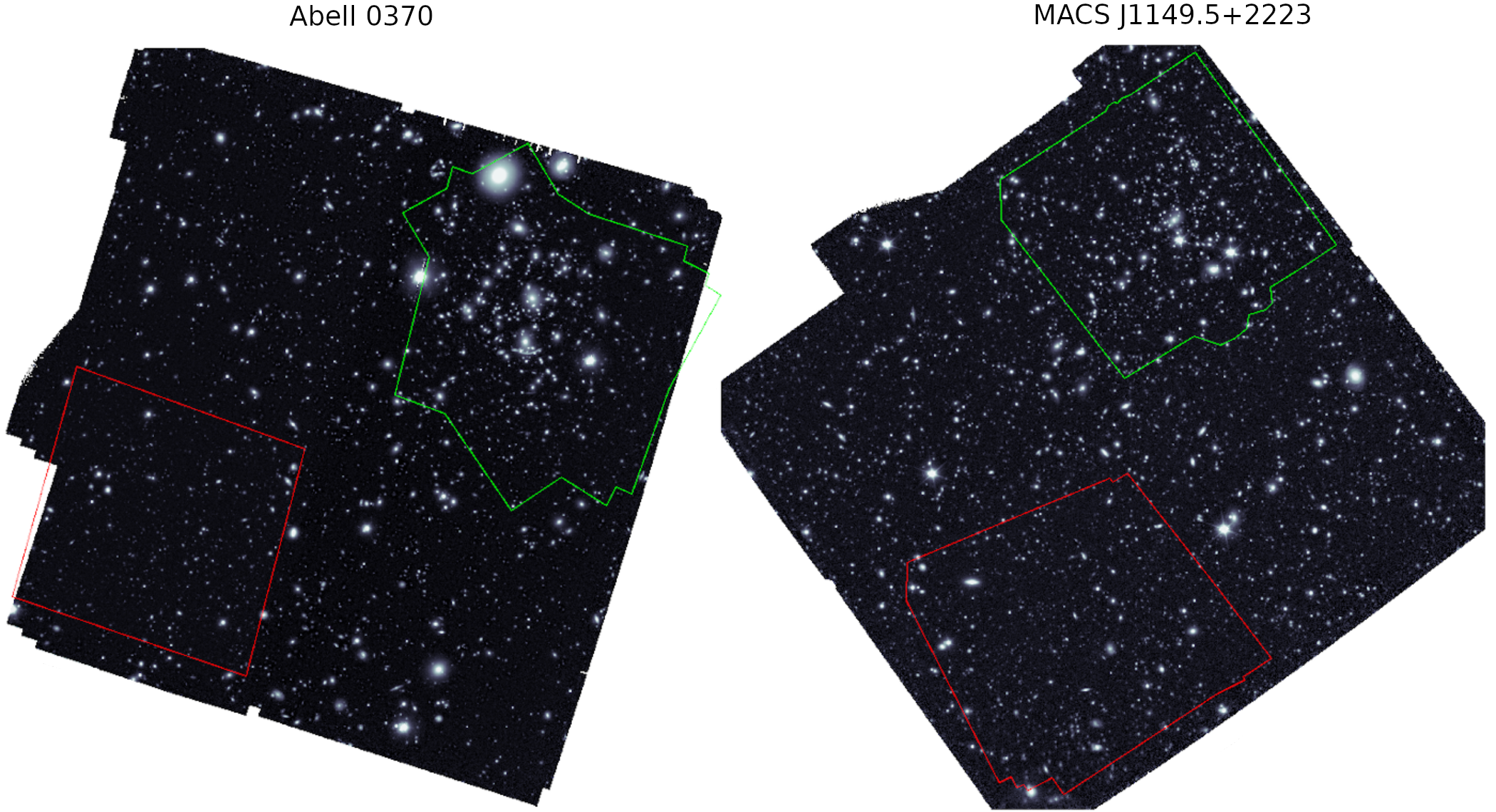}
	\caption{Observational field-of-view comparison. SHARDS F913W25 observations of the A0370 (left) and M1149 (right) fields covering roughly $\sim$70\arcmin$^2$ each. Overlaid in green and red are the composite detection image FOVs for the Hubble space telescope cluster and parallel fields respectively (note that not all of the area within the green and red lines is observed by both WFC3 and ACS). The Hubble data we use comes from  from a combination of the F814W, F105W, F125W, F140W and F160W bands, we refer the reader to \citet{Shipley2018} and \citet{Lotz2017} for more information.  Essentially, the Hubble field of view is only about 50\% of that of the SHARDS imaging field of view.}
	\label{fig:obsoverlay}
\end{figure*}

For differential flux measurements and the selection of emission line galaxies, we require deep broad-band (BB) multi-wavelength data covering the same wavelengths as the medium-band filters. For this reason we make use of catalogues and imaging data made available as part of the Hubble Frontier Fields (HFF) Deep Space project \citep{Shipley2018}. These data combine up to 17 ACS/WFC3 filters with ultra-deep Ks band imaging, and Spitzer-IRAC, when available. The HFF-DeepSpace dataset also includes calibrated catalogues, photometric redshifts, lensing magnification factors as well as original imaging data, models and calibration information, providing an ideal ancillary dataset for our candidate selection. We note here that HFF-DeepSpace observations cover only a fraction of the $\sim$70 square arcminute area surveyed by SHARDS-FF for each cluster. Considering both the cluster and parallel field, this constitutes roughly $\sim$35\%~and 38\% of the total SHARDS-FF coverage for A0370 and M1149 respectively. In Figure~\ref{fig:obsoverlay} we show the HFF-DeepSpace detection image FOV for both clusters and their parallel fields overlaid on SHARDS F913W data.  

\subsection{Photometric calibration}
\label{sec:calib}

Candidate line emitter selection requires accurate photometric calibration between all observations. This presents a significant challenge when combining ground and space based data due to the differences is spatial resolution, PSFs etc. In order to optimally combine the data it is necessary to avoid the degradation of the HST imaging to the lower resolutions of the ground based SHARDS observations, while conversely not artificially up-scaling the low resolution images, likely to provide unreliable flux measurements. To circumvent these issues, we opt to calibrate the medium-band flux measurements following the same procedures used to construct the HST catalogues \citep{Shipley2018} and examine this methodology in the sections below.        

\subsubsection{Total Flux Measurements}

All observations are first matched to the PSF of the SHARDS F883W filter ($\sim$1.0\arcsec).  This is done by deriving convolution kernels for each band using the PSFs provided as part of both the SHARDS and HFF-DeepSpace datasets. For all HFF-DeepSpace imaging we use 2\arcsec~apertures to recalculate photometry in SExtractor's dual image mode, utilising the deep detection images from the HFF-DeepSpace data \citep[see][section 3.3]{Shipley2018}. We perform aperture photometry for SHARDS bands individually, constructing a final catalogue by matching sources detected in the SHARDS filters to the HFF-DeepSpace IDs.  

To correct for flux falling outside of the 2\arcsec~apertures, we derive total flux values. Aperture photometry is adjusted by applying a correction factor, derived on a source-by-source basis from a reference band (typically F160W). Firstly, for each galaxy the reference band total flux ($f_{ref,tot}$) is calculated from the SExtractor \citep{Bertin1996} AUTO flux using a growth curve, in combination with the Kron radius \citep[for full details see][]{Shipley2018}. A conversion factor is then calculated from the ratio of the reference bands total flux to the corresponding 2\arcsec~aperture flux ($f_{ref}(r)$). Following the equation:

\begin{equation}
    f_{i,tot} = f_{i}(r)\frac{f_{ref,tot}}{f_{ref}(r)}
    \label{eq:fluxconv}
\end{equation}

\noindent where r is the aperture radius, total fluxes for all other bands ($f_{i,tot}$) can be derived based on the measure aperture flux, $f_{i}(r)$.

It should be noted here that this method of producing our catalogues only provides SHARDS medium-band flux measurements for sources which are also found in the detection HST images (i.e. sources that are emitting in the continuum).   Below we describe how we apply further flux corrections which are required to robustly combine the medium-band imaging with HFF-DeepSpace data.  

Typically, when investigating galaxies within the fields of massive clusters such as those in this study, magnifications effects resulting from strong gravitational lensing should be considered. Our H$\alpha$ sample, the focus of this paper and which is described in \S 3.3, is however situated at, or below the cluster redshifts such that corrections for these lensing effects are unnecessary.

In order to obtain consistent measurements between broad- and medium-band photometry, we apply further flux corrections. We correct for Galactic extinction, taking values given by the NASA Extragalactic Database extinction law calculator\footnote{http://ned.ipac.caltech.edu/help/extinction\_law\_calc.html} for the centre of each field and filter \citep[for a full breakdown of extinctions used for HFF-DeepSpace filters, see Table 5 of][]{Shipley2018}. Further, flux values are normalised to a zero-point of 25.   

\subsubsection{Geometric effects}

Geometric effects due to incidence angles of the GTC/OSIRIS light beam result in spatially varying effective central wavelengths (CWL) within each filter. This must be corrected for.   Sources within each SHARDS observation are detected with similar, but not identical, effective central wavelengths depending on their location within the image. These effects have been calibrated for previously in \citet{Perez2013}.  We also need to included this correction in this work. The variation of the filter's central wavelengths will not only affect the SED and SPS fitting, but also calculations of flux densities, this in turn will induce some shift in the calculated equivalent width and excess significance parameters used for the selection of candidate emitters. Thus, accurate calibration of geometric effects need to be undertaken.

Here, we present a summary of this calibration procedure which is described in detail in \citet{Perez2013} (Section~3.3). The central wavelength calibration is performed through day-time imaging and laboratory obtained spectroscopic data. Using a pinhole mask, spectra are obtained covering the entire FOV, from which the transmission curve is measured. The shape and width of the curve remain relatively constant while the central wavelength shows systematic variations around the optical axis (to the left of the FOV). Central wavelengths are fit with a function that depends on the distance from the optical axis squared ($r^2$). Given the source location and the position of the optical axis, the central wavelength can be computed following the equation:

\begin{equation}
	CWL\left(X,Y\right)=A+B\times\left[\left(X-X_0\right)^2+\left(Y-Y_0\right)^2\right]
    \label{eq:cwl}
\end{equation}

\noindent where $X,~Y$ $X_0,~Y_0$ are the pixel locations of the sources, and the optical axis respectively. The values $A$ and $B$ are the fitting coefficients, and along with the values of $X_0$ and $Y_0$ are filter dependent and do not vary with time. The values of these are presented in Table~1 of \citet{Perez2013}. Calibrations were tested for repeatability and found to provide robust measurements over different nights.


\subsection{Updated Photometric Redshifts and Stellar Population Parameters}
\label{sec:photz}

We utilise our photometric catalogues combining broad-band data from HFF-DeepSpace with our calibrated medium-band photometry to obtain high quality photometric redshifts and stellar population parameters for all galaxies within both of the clusters, and their parallel fields. The photometric fitting codes EAZY \citep{Brammer2008} and FAST \citep{Kriek2009} are used for this as default parameter files are provided in the HFF-DeepSpace dataset. This allows for the most robust comparison to existing catalogue values. For a detailed description of parameter selection and verification see Sections~5.2 and 5.4 of \citet{Shipley2018}.

\subsubsection{Photometric Redshifts} \label{sec:eazy}

Photometric redshifts are computed with the SED fitting code EAZY\footnote{https://github.com/gbrammer/eazy-photoz/} \citep{Brammer2008}. We use a linear combination of 12 galaxy templates based on Flexible Stellar Population Synthesis (FSPS) models \citep{Conroy2009,Conroy2010} and trained on the UltraVISTA photometric catalogues \citep{Muzzin2013}. We implement a redshift prior based on F160W apparent magnitudes and use a scaled default template error function. 

\begin{figure}
	\includegraphics{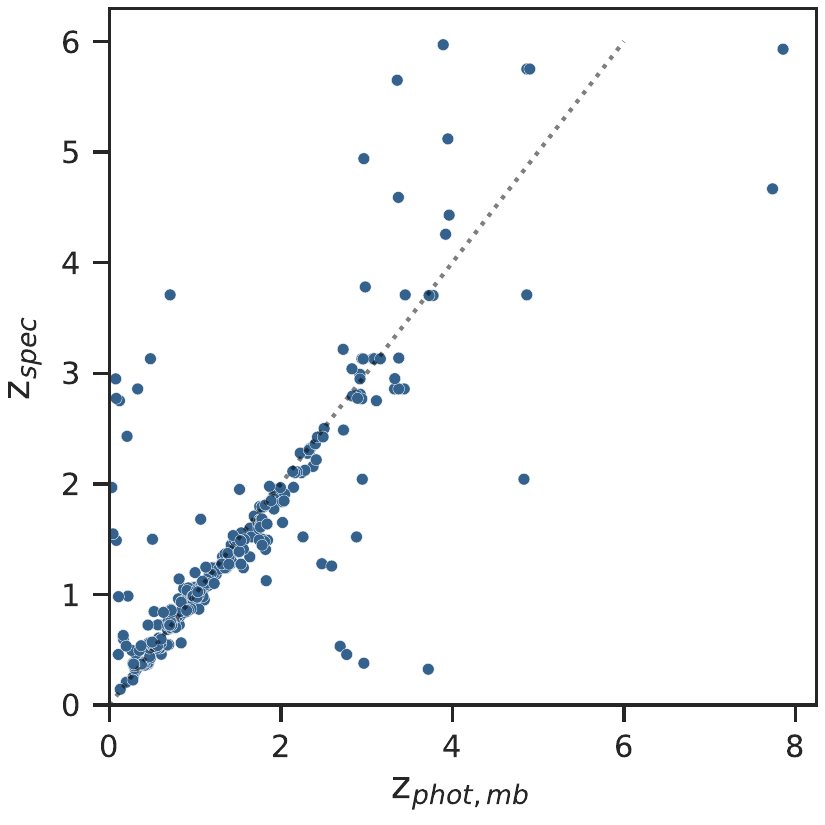}
	\caption{Comparison of the new photometric redshifts we fit in this study ($z_{phot,mb}$) to the 502 available spectroscopic measurements. Dotted line shows the one-to-one relation. We find an average scatter of 0.073 and a catastrophic outlier rate of 9\% (see text).}
	\label{fig:specz}
\end{figure}

Due to the variations in the effective central wavelength of the SHARDS filters, EAZY must be run on a source-by-source basis. We employ a custom python code to automate this process and incorporate the object specific filter response curves. In summary; for each object in our catalogue, the code uses a pre-defined set of EAZY parameters in combination with our photometric data to create a set of temporary input files. The SHARDS filter response curves are shifted to the effective central wavelength before EAZY is run on the single object. Once results are obtained for each object, the code then compiles the data into a single catalogue. Spectroscopic redshifts are used when available, otherwise the peak of the photometric redshift distribution (\textit{z\_peak}) is taken as the redshift. We compare our updated results with spectroscopic and photometric redshifts from the HFF-DeepSpace catalogues in Figures~\ref{fig:specz} and \ref{fig:comp} respectively.

We fit photometric redshifts for a total of 24,183 galaxies over the four fields, 502 of which have accurate spectroscopic redshift measurements available from a variety of sources. For these 502 objects, we compare the results of our photometric redshift fits to the spectroscopic measurements,  findind an average scatter of the residual of $|\Delta z|/(1+z) = 0.073$, and using the common definition of catastrophic outlier, $|\Delta z| > 0.15(1+z_{spec})$ \citep[e.g.][]{Ilbert2009,Dahlen2013}, we find a rate of 9\%. We show this comparison in Figure~\ref{fig:specz}.

\subsubsection{Stellar Population Parameters}
\label{sec:sps}

Stellar population parameters such as stellar mass, star formation rates and ages are estimated for our sample of galaxies with FAST\footnote{http://w.astro.berkeley.edu/~mariska/FAST.html} \citep{Kriek2009}. For input parameters we again follow the methodology of \citet{Shipley2018}; employing a Chabrier IMF \citep{Chabrier2003}, solar metallicity, minimum star-formation age of 10 Myr, Calzetti dust attenuation law \citep{Calzetti2000}, $0 < A_{V} < 6$ mag and exponentially declining star formation histories with a minimum e-folding time of $\log_{10}(\tau/yr) = 7$. We use a Bruzal \& Charlot SPS model library \citep{Bruzual2003} which includes random bursts (of duration between $3x10^{7} - 3x10^{8}$ yr) superimposed onto models of continuous star formation histories (SFHs) with a probability that 50\% of galaxies have experienced a burst of the past 2 Gyr.  We use a solar metallicity here as this is what has been used previously in the HFF-DeepSpace catalogue. Furthermore, the use of solar metallicity only affects the stellar mass measurements, and if our systems are significantly less than solar this would produce stellar masses which are higher than they should be by +0.3 dex, which is not enough to alter any of our conclusions.  A higher solar metallicty may also affect our dust measurements, as the galaxies would be slightly redder when fit and thus would require less dust. FAST models do not currently include emission lines, however their effect on stellar mass estimates has been shown to be small \citep[e.g.][]{Banerji2013}, which would especially be the case for lower mass galaxies.

\begin{figure*}
	\includegraphics{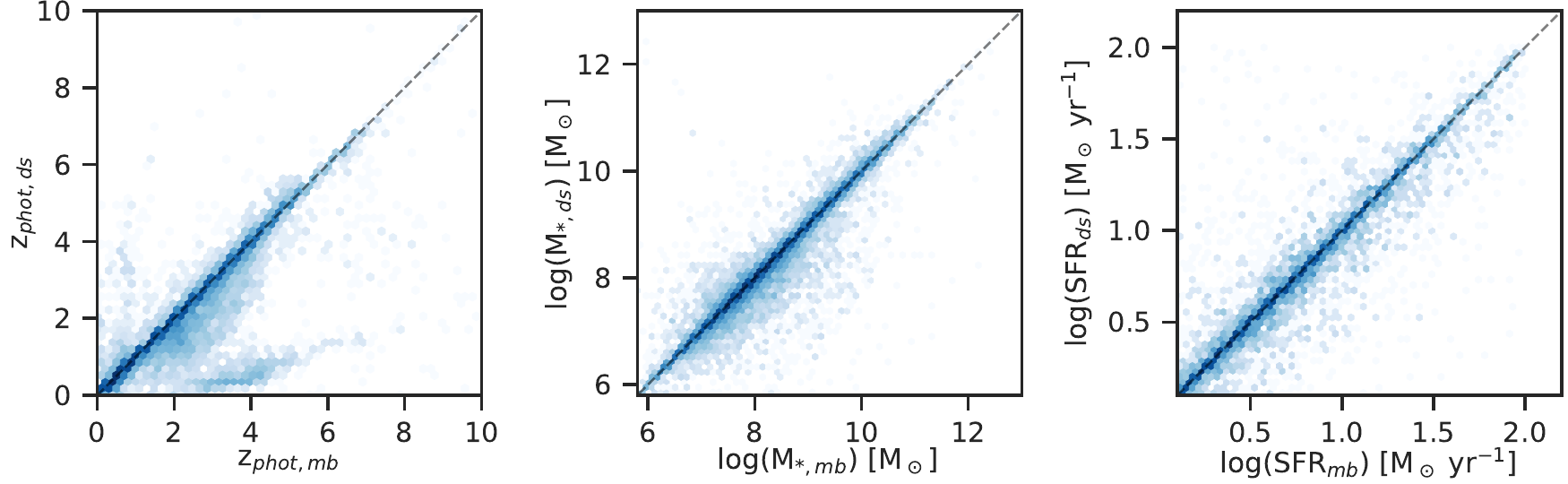}
	\caption{Photometric redshift and stellar population parameter for all objects in both the cluster and their corresponding parallel fields obtained in this work, compared to HFF-DeepSpace catalogue values. Left panels shows a direct comparison of photometric redshifts estimated with EAZY from HFF-DeepSpace catalogues and this work (HFF-DeepSpace + SHARDS filters). The grey dotted line represents a one-to-one ratio. Middle and right panels are similar, but for the FAST estimates of stellar mass and star formation rate respectively. In all of the plots, the subscript `mb' (medium-band) represents estimates obtained in this work, while the subscript `ds' (DeepSpace) denotes the values obtained from the HFF-DeepSpace catalogues.}
	\label{fig:comp}
\end{figure*}

In order to obtain the most accurate stellar population parameters, we again shift the SHARDS filter response curves on an object to object basis to account for variations in central wavelength and use our updated photometric redshift estimates (see Section~\ref{sec:eazy}). While the inclusion of the SHARDS medium-band photometry is likely to improve estimates of these parameters, only mass-to-light ratios are well constrained due to their dependence on rest-frame optical colours which are covered by the HFF-DeepSpace photometry. We show in the central, and right panels of Figure~\ref{fig:comp} that our results on measuring these features with our broad band and narrow band imaging are in good agreement with HFF-DeepSpace parameters.

\section{Emission Line Galaxy Selection}
\label{sec:cand}

In order to select emission line objects via differential flux measurements, we require overlapping wavelength coverage of both the medium and broad-band imaging. We utilise F606W and F105W broad-band imaging to match the F517W and F941W medium-band filters respectively, while F814W is matched to both the F823W and F913W bands (see Figure~\ref{fig:filters}). We note here that the sky area covered by the F105W band is less than half the size of that covered by the F814W imaging (only $\sim$42\% and $\sim$34\% the size of the F814W science area for the cluster and parallel fields respectively). For the robust selection of emission line objects we employ a two parameter selection criterion based on emission line equivalent width and having an excess significance, as is well established in previous studies \citep[e.g.][]{Matthee2015,Santos2016,Sobral2017,arrabal2018}. These criterion assures that the objects selected show a real colour excess, and not an excess due to random scatter or measurement uncertainties. 

\subsection{Broad-band Continuum}

\begin{figure}
	\includegraphics{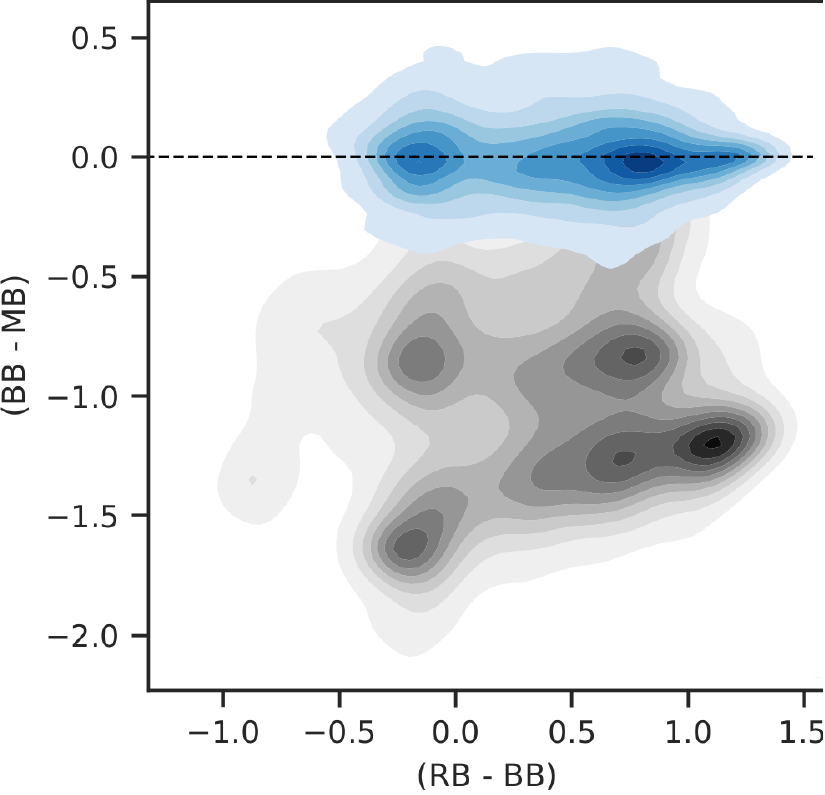}
	\caption{Colour-colour plot showing broad-band continuum magnitude corrections for all objects in both Frontier Field clusters and their corresponding parallel fields. We show object colours before (grey contours) and after (blue contours) corrections have been applied to achieve a mean-zero BB-MB colour.}
	\label{fig:bbfix}
\end{figure}

Before applying the two parameter selection criterion, it is important to note that the medium-band filter profiles are not centrally aligned with the corresponding broad-band filters. Thus, sources with significant intrinsic colour in the continuum can complicate not only the selection of emission line objects, but also result in over or underestimation of line fluxes. In order to correct for this we derive an effective broad-band magnitude (called BB') to account for the slope in the continuum and achieve a mean zero (BB-MB) colour \citep{Sobral2012,Sobral2013}. To obtain the most accurate estimates we first remove all point sources from the catalogue using the \textit{star\_flag} identifier. A further cut is then performed to select sources within 2$\sigma$ of the scatter around the median of the broad-band colour. We then perform a linear fit which is used to correct the initial broad-band magnitude following the form:

\begin{equation}
	\left(BB'-MB\right)=\left(BB-MB\right)-M(RB-BB)+C
    \label{eq:bbcor}
\end{equation}

\noindent where $RB$ is the reference band (taken as the closest neighbouring broad-band filter), and M and C are the coefficients of the linear fit. For sources in which one of the broad-band colours is not available, the median correction is applied. We show (BB-MB) colours before and after corrections have been applied in Figure~\ref{fig:bbfix}.  The net effect of this is to reduce our number of candidates significantly.  We take a conservative apporach here so as to remove as much as possible contamination and to retain a population of certain line emitters.

\subsection{Line Emitter Selection}
\label{sec:selection}

To carry out a selection for our emission line candidates, first we make a check to determine if the medium-band excess is high enough to be considered an emission line object. This is done by setting a lower limit on equivalent width (EW). The observed equivalent width is the ratio between the flux of an emission line (medium-band) and the continuum (broad-band). We follow the basic procedure used in previous searches and specify a traditional excess criteria of rest-frame EW of 25\AA~\citep[e.g.][]{Ouchi2008}. Previous studies \citep{Ouchi2010} have found that higher EW cuts ($>$25\AA) help to minimise contamination from low redshift interlopers, while \citet{Sobral2017} are able to implement much lower cuts through the use of narrower filters. Due to the limited sample size and low expected EW at our modest redshifts we use the lower limit of 25\AA. To obtain an EW from the observed medium-band excess, magnitudes ($m_i$) is converted to flux densities ($f_i$) for each filter ($i$) via the equation: 

\begin{equation}
	f_i=\frac{c}{\lambda_{i,centre}^2}10^{-0.4(m_i+48.6)}
    \label{eq:flux}
\end{equation}

\noindent where $c$ is the speed of light and $\lambda_{i,centre}$ is the filter's central wavelength (note that for the medium-band filters, this is taken as the corresponding broad-band central wavelength as the continuum colour corrections of Equation~\ref{eq:bbcor} have been applied). Line fluxes and equivalent widths are then computed using:

\begin{equation}
	f_{line} = \Delta\lambda_{MB} \left(\frac{f_{MB}-f_{BB}}{1-\frac{\Delta\lambda_{MB}}{\Delta\lambda_{BB}}}\right)
    \label{eq:fline}
\end{equation}

\begin{equation}
	EW = \Delta\lambda_{MB}\left(\frac{f_{MB}-f_{BB}}{f_{BB}-f_{MB}\frac{\Delta\lambda_{MB}}{\Delta\lambda_{BB}}}\right)
    \label{eq:ew}
\end{equation}

\noindent respectively, where $\Delta\lambda_{MB}$ and $\Delta\lambda_{BB}$ are the widths of the medium- and broad-band filters respectively. Here, $f_{MB}$ and $f_{BB}$ are the flux densities calculated via Eq.~\ref{eq:flux}. Depending on the specific filters used, this equation breaks down at certain MB excess when $\Delta\lambda_{MB}/\Delta\lambda_{BB} \to f_{BB}/f_{MB}$ such that the denominator tends to 0. When this occurs we set the EW of the sources to > 1500\AA~as this can also be helpful to identify sources that may not be real, or where problems may have arisen.

The second parameter, the excess significance \citep[$\Sigma$, e.g.][]{Bunker1995} is used to quantify the real flux excess compared to an excess due to random scatter. The excess significance can be written as \citep{Sobral2013}: 
\begin{equation}
	\Sigma=\frac{1-10^{-0.4(BB-MB)}}{10^{-0.4(ZP-MB)}\sqrt{\pi r_{ap}^2(\sigma_{px,BB}^2-\sigma_{px,MB}^2))}}
    \label{eq:sig}
\end{equation}
where $MB$ and $BB$ are the broad- and medium-band magnitudes respectively. Here, $ZP$ represents the normalised zero point (25) and $r_{ap}$ is the aperture radius in pixels. The root-mean-square (rms) of the background pixel values, $\sigma_{px}$, are estimated by randomly placing empty apertures across the respective images. A selection criteria of $\Sigma$ > 3 is used to classify sources as potential line emitters \citep{Sobral2013}.

\begin{figure*}
	\includegraphics{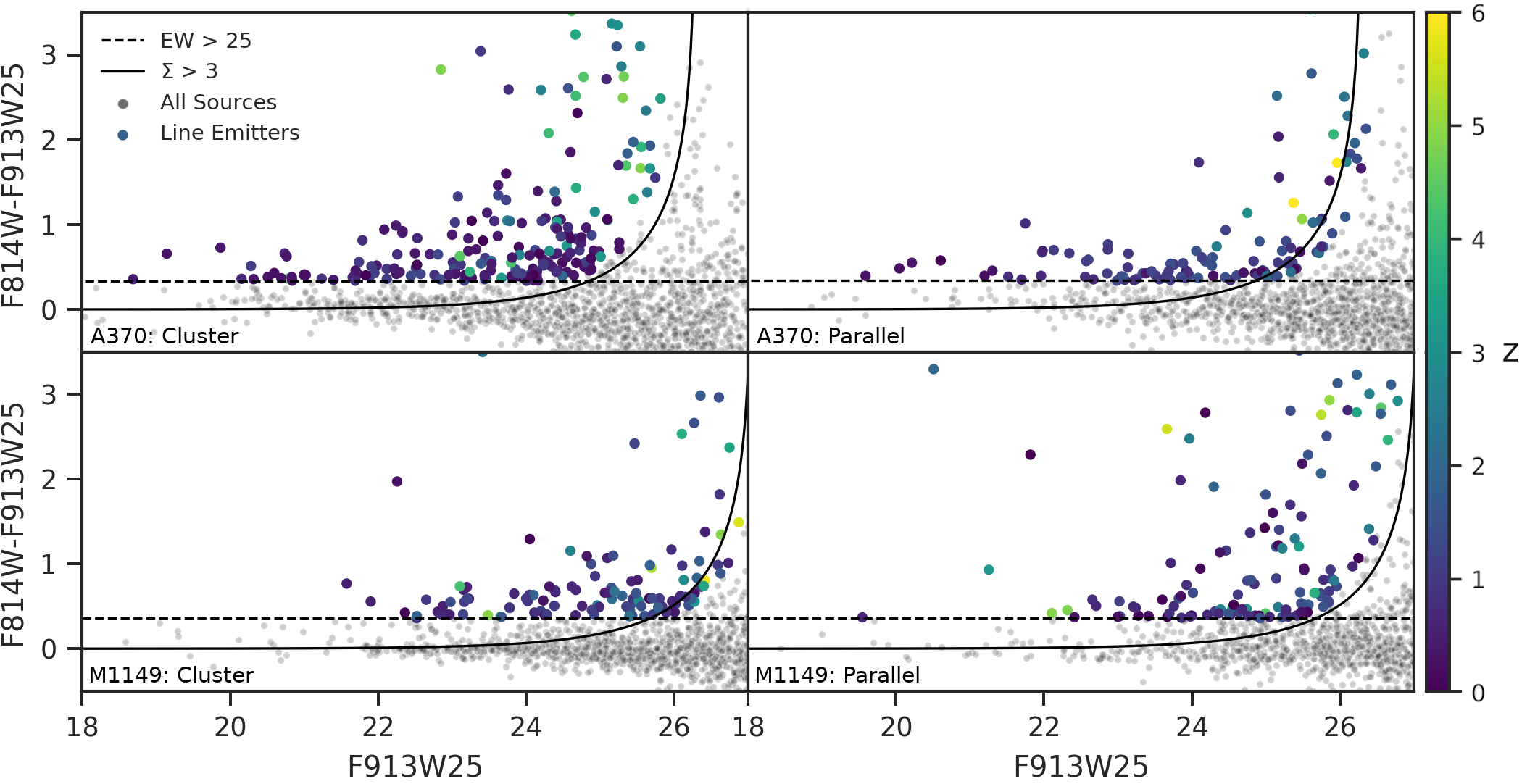}
	\caption{Medium-band excess as a function of F913W medium-band magnitude for the A0370 and M1149 clusters (top left and bottom left panels respectively) and their parallel fields (top right and bottom right panels). The horizontal dashed lines represents a rest-frame EW cut of 25\AA~while solid lines show the average 3.0$\Sigma$ colour significance. All objects are shown as grey points while galaxies that meet the selection criterion are coloured by photometric redshift.}
	\label{fig:selection}
\end{figure*}

\begin{figure*}
	\includegraphics{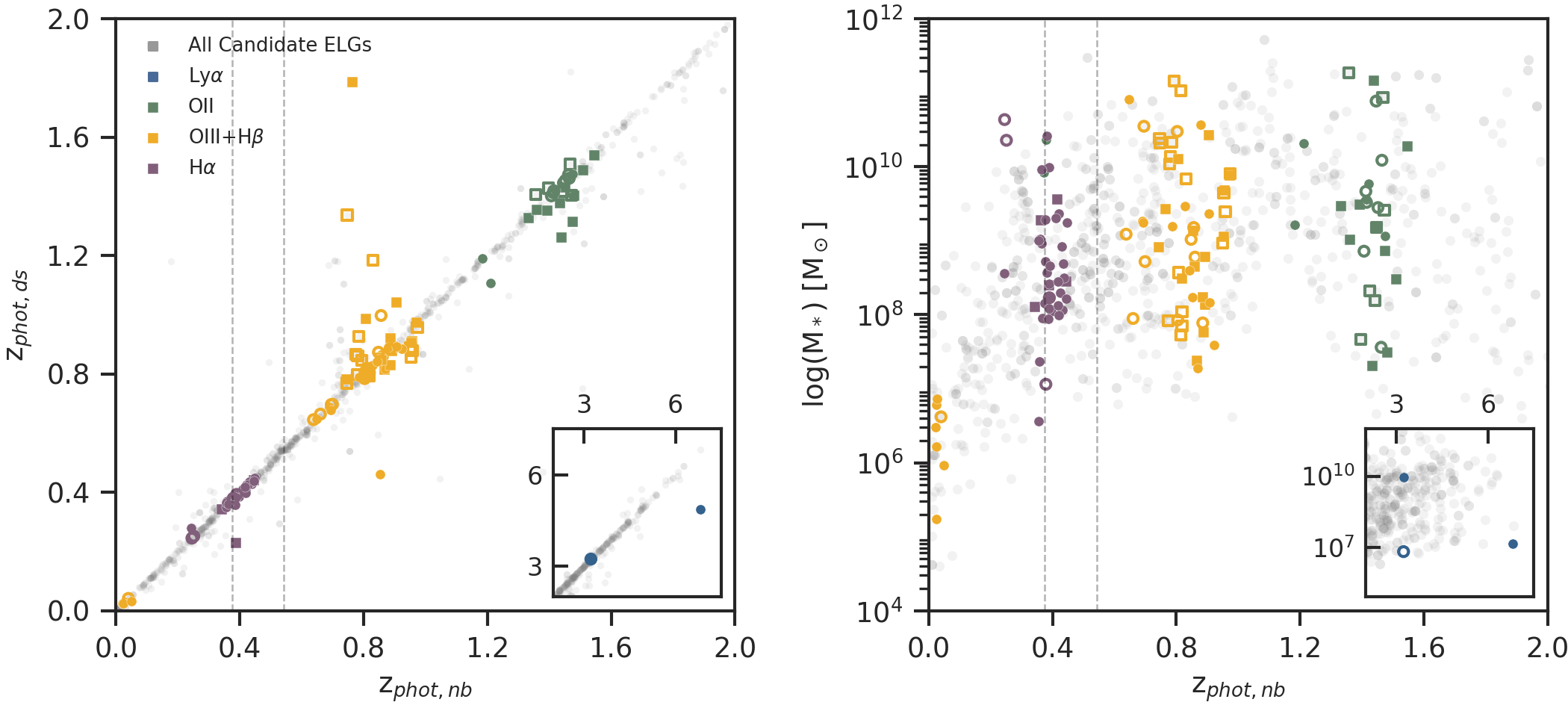}
	\caption{Left panel: redshift comparison of emission line selected candidates with ($z_{phot,mb}$), and without ($z_{phot,ds}$) SHARDS medium-band filters. We show a few example emission lines based on the detected medium-band filter and photometric redshift, other emission line candidates are shown as grey points. Candidates selected from the A0370 and M1149 fields are shown as circles and squares respectively, while filled and empty points represent the cluster and parallel fields. The redshift of the clusters are shown by dashed vertical lines at 0.375 (A0370) and 0.543 (M1149). The inset shows the same again from $2 < z < 7.5$, illustrating the detection of 3 Lyman-$\alpha$ emitters. Right panel: Stellar mass as a function of photometric redshift for all candidate objects. Coloured points correspond to the same emission line selected candidates as in the left panel.}
	\label{fig:candidates}
\end{figure*}

We show the selection of potential line emitters in Figure~\ref{fig:selection}. From $\sim$25,000 detected objects across the four fields (two clusters and their respective parallels), 666 candidates are found in the cluster fields with a further 432 in the parallels, providing a total of 1,098 candidate line emitters. Through the use of our photometric redshift estimates with the detection filter throughput information, it is possible to assign suspected emissions lines to objects within our sample. This method uses our above criteria and removes a significant number of candidate sources. Ultimately, we are left with only a small fraction which are reliable emitters.  One of the main reasons for this is that we are fairly conservative about the line matches with wavelength.  This is one of he deepest narrow band searches yet, and it is also possible that many of the lines are from rare ionization lines that are not the common lines often seen in narrow-band searches, such as H$\alpha$, \ion{O}{II} and \ion{O}{III}.  The comparison of our numbers agrees with the expected uncertainties with previous work such as \citet{Sobral2013}. For example; we find 42 robust H$\alpha$ candidates as well as 33 \ion{O}{II}, 63 \ion{O}{III}$+$H$\beta$ and 3 suspected Ly$\alpha$ line emitters. We also find 42 candidate emission line galaxies between redshifts $2 < z < 3.5$ which we use to investigate escape fractions in Paper II (Griffiths et al. 2021, submitted). In Figure~\ref{fig:candidates} we show the updated photometric redshifts for our candidate emission line galaxies as a function of HFF-DeepSpace photo-z (i.e. without SHARDS medium-band filters) and updated stellar mass (see Section~\ref{sec:sps}), highlighting a few example emission lines only.
 
\subsection{H$\alpha$ Emitter Selection}
\label{sec:ha}

We select H$\alpha$ emitters, which are the focus of the rest of this paper, from our emission line sample based on a combination of the detected filters and updated photometric redshifts. We utilise filter throughput information for all medium- and broad-band filter combinations to determine the redshift range in which the detected emission line coincides with H$\alpha$. When a galaxy is selected as an emission line candidate via the two parameter selection criterion (as described in Section~\ref{sec:cand}) and its photometric redshift coincides with the H$\alpha$ emission line for the detection filter combination, the object is marked as a H$\alpha$ emitter. We find a total of 42 objects meeting this criterion over the four fields (the two clusters and their corresponding parallel fields), and summarise the redshift range for all filter combinations in Table~\ref{tab:haz}.  Other line emitters will be the focus of future papers.

Given the luminosity functions of \citet{Sobral2013} at $z=0.4$, we would expect to find roughly 4 H$\alpha$ emitters within our fields. This is consistent with our results if the overdensity in the A0370 cluster field can be attributed to the sources being cluster members. This is further supported by the work of \citet{Stroe2017} in which 9 narrow band filters are used to select over 3000 H$\alpha$ emitters from 19 galaxy clusters and their large scale environments (beyond 2 Mpc from cluster centre). Their results indicate that cluster fields are overdense in H$\alpha$ emitters, with the luminosity function showing a strong dependence on the dynamic state of the cluster. Considering recent simulations suggesting A0370 is a merging system \citep{Molnar2020} along with the volume probed by the SHARDS data, our results are consistent with the work of \citet{Stroe2017} finding similar number densities of H$\alpha$ emitters, considering our deeper depth. This also confirms our approach towards finding these emitters is sound and consistent with previous work.

We also visually inspect all objects which are within a crowded environment or contain contaminating flux from neighbouring sources. In total we find four candidates which can not be successfully isolated. These objects are excluded from our analysis as the overlapping light profiles have the potential to result in unreliable photometry, as well as systematic shifts in our morphological and structural measurements. 

\begin{figure}
	\includegraphics[width=\columnwidth]{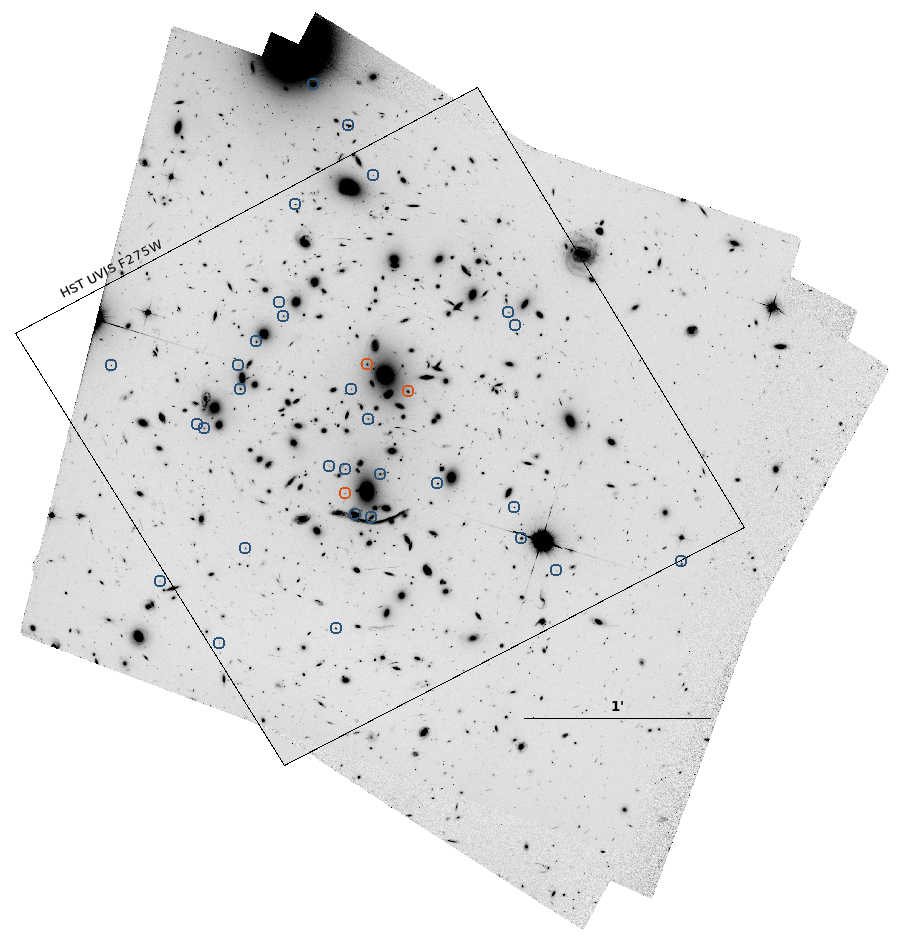}	
	\caption{Hubble Space Telescope F814W image of the A0370 cluster field with locations of candidate H$\alpha$ emission line sources denoted by open blue circles. Orange circles mark the location of objects found to be undergoing a recent burst of star formation following the analysis in Section~\ref{sec:bursts}. The HST/UVIS F275W FOV is displayed by the black contour. } 
	\label{fig:locations}
\end{figure}

A total of 33 of the remaining 38 candidate galaxies are selected from the A0370 field.  This is expected as the redshift of the A0370 cluster ($z = 0.38$) corresponds to that of our detection range such that we are able to select both foreground galaxies and cluster members. We show the distribution of our H$\alpha$ candidates within the A0370 field in Figure~\ref{fig:locations}. However, as M1149 is at the slightly higher redshift of $z=0.54$ we are unable to probe the cluster population with our permitted filter combinations.      This is a strong indication that our H$\alpha$ emitters are located mostly within dense environments.  

We note here that this process is undertaken for all chosen emission lines, however a full analysis of all these emitters is beyond the scope of this paper and will be discussed in future publications. 

\begin{table}
\caption{Detection filter redshift range and cluster candidate counts of H$\alpha$ emitters for the MACS J1149.5+2223 (M1149) and Abell 370 (A0370) fields.}
\label{tab:haz}
\centering
\begin{tabular}{ccccc}
  \hline
  MB  &  BB  &  Redshift Range  & \# M1149 & \# A0370 \\
  \hline
  F823W  &  F814W  &  0.24 $<$ z $<$ 0.27 & - & 3 \\
  F883W  &  F814W  &  0.32 $<$ z $<$ 0.37 & 2 & - \\
  F913W  &  F814W  &  0.37 $<$ z $<$ 0.41 & 1 & 19 \\
  F941W  &  F105W  &  0.41 $<$ z $<$ 0.46 & 2 & 11 \\
  \hline
 \end{tabular}
\end{table}

\subsubsection{Correction for [N\small\textsc{II}] line contamination}

Calculations of H$\alpha$ emission line fluxes and equivalent widths must take into consideration the contribution of flux from neighbouring [\ion{N}{II}] lines at $\lambda_{rest}$ = 6548, and 6583\AA. As these lines are situated in close proximity to H$\alpha$, their flux is typically included in photometric measurements, acting to artificially increase line fluxes and EWs measured. To account for this, we implement a correction based on the work of \citet{Villar2008} where it was found that the fractional contribution of the [\ion{N}{II}] flux decreases with increasing EW due to lower metallicities. The flux ratio, $F_{\rm{N_{II}}} / F_{\rm{H_{\alpha}}}$ and equivalent width, EW(H$\alpha$+[\ion{N}{II}]) are related by:

\begin{equation}
	\log\left({F_{\rm{N_{II}}}/F_{\rm{H_{\alpha}}}}\right) = -5.78 + 7.63x - 3.37x^2 + 0.42x^3
    \label{eq:ffix}
\end{equation}

\noindent where the value "x" is log(EW(H$\alpha$+[\ion{N}{II}])). We use this relation to correct all H$\alpha$ fluxes in this study whereby we a  find a median correction of $\sim$25\%.   The range of masses for our sample overlap with the range of masses for which this criteria was measured and determined.  

\subsubsection{Extinction correction}
\label{sec:extcor}

The H$\alpha$ emission line is much less affected by dust obscuration than other common lines such as Lyman-$\alpha$ or [\ion{O}{II}], however dust effects are non-negligible and should be accounted for.  This is particularly the case when deriving star formation rates. When spectroscopy is available for individual sources, the amount of extinction can be estimated by comparing the intrinsic Balmer decrement with that observed. As spectroscopy of each individual sources is often unfeasible, dust extinctions can also be estimated via the comparison of H$\alpha$ and far-infrared based SFRs \citep{Ibar2013,Koyama2013}.


There are many ways in which to account for dust extinction in our sample, some of which we explore below, and throughout the paper. Some studies employ a simple approach of applying a constant value of dust extinction throughout a galaxy. As determined by \citet{Kennicutt1992} a 1 mag extinction of H$\alpha$ can be adopted, and is often a good approximation for the true extinction \citep[e.g.][]{Ly2007,Geach2008,Sobral2009}.  However, this extinction has been found to be stellar mass dependent in the local universe \citep{Garn2010b}, in clusters \citep{Sobral2016}, and at high redshift \citep{Sobral2012}, so it must be used with care.

However, as SEDs are readily available for our galaxy candidates, we estimate dust attenuation based on stellar E(B-V) values derived during fitting (see Section~\ref{sec:photz}) and the assumption of $A_{\rm{v,stars}} = 0.44A_{\rm{v,gas}}$ found in local starburst galaxies \citep{Calzetti2000}.  This is a major assumption we are making, and we later investigate how our results would differ if we were to assume other relationships, such as equality between the stellar and gaseous components.  We find all galaxies have a dust content such that 0 $<$ E(B-V) $<$ 0.7 with mean, median and dispersion values for our dust corrections are 0.16, 0.06, and 0.20, respectively.  

We also investigate the situation whereby we make the assumption that
$A_{\rm{v,stars}} = A_{\rm{v,gas}}$, which in some higher redshift situations is more likely the case \citep[e.g.,][]{reddy2015}.   We describe in detail how this assumption would change our results when we discuss the effects of star formation.  The net effect is that the measured star formation rate in the UV would be $\sim1.85$ times lower with this equality in the dust extinction. This produces a change of $-0.26$ dex when plotting features in log space. As we discuss later, this affects in no significant way the conclusions we draw from this study. This shows that these galaxies are overall not very dusty systems, which fits in with them typically being low-mass galaxies.

\subsubsection{Star formation rates}
\label{sec:sfrs}

A galaxy's star formation activity changes throughout its lifetime due to various physical processes such as major/minor mergers, gas accretion, and supernovae feedback. These fluctuations all contribute to the luminosity of the H$\alpha$ emission line, which is sensitive to instantaneous star formation on timescales of the order $\sim$10 Myr \citep[e.g.][]{Kennicutt1998}. The burstiness of star formation activity can be investigated through the direct comparison of H$\alpha$ derived SFRs with those estimated via UV continuum emission, which provides time averaged SFRs over a much longer $\sim$100 Myr period \citep[e.g.][]{Kennicutt1998}. It is also the case that the H$\alpha$ equivalent width (EW) is a good indicator for the specific star formation rate \citep{khostovan2020}. 

As the UV continuum traces star formation over a wider range of stellar ages than that of H$\alpha$, our calibrations must trace a galaxy's SFR history for at least the last 100 Myr.  As such, we derive UV star formation rates from rest-frame 1600\AA~emission (L$_{\rm{UV}}$) for our H$\alpha$ sample.  The UV flux corresponds to the F275W band, which is provided in the HFF-DeepSpace dataset for both clusters.  The luminosities for these galaxies are first corrected for reddening using the stellar E(B-V) values derived from SED fitting before SFRs are obtained from the \citet{Kennicutt1998} relation adapted from a Salpeter, to \citet{Chabrier2003} IMF following \citet{Muzzin2010}:

\begin{equation}
	\rm{SFR(M}_{\odot} yr^{-1}) = 0.8 \times 10^{-28} L_{\rm{UV}} (erg~s^{-1}~Hz^{-1})
    \label{eq:uvsfr}
\end{equation}

\noindent To provide a consistent set of calibrations, we convert corrected H$\alpha$ luminosities into star formation rates using the standard calibration of \citet{Kennicutt1998}, again modified to a \citet{Chabrier2003} IMF following \citet{Muzzin2010}:

\begin{equation}
	\rm{SFR(M}_{\odot} yr^{-1}) = 4.5 \times 10^{-42} L_{H_{\alpha}} (erg~s^{-1})
    \label{eq:hasfr}
\end{equation}

\noindent assuming Case B recombination at T$_e = 10^4$K and continuous star formation. All measurements are based on line luminosities in which the [\ion{N}{II}] contamination has been accounted for and dust attenuation is assumed to follow $A_{\rm{v,stars}} = 0.44A_{\rm{v,gas}}$, as discussed in Section~\ref{sec:extcor}. While this calibration assumes continuous star formation on timescales of over 100 Myr, it is relatively robust to variations in recent star formation history (if it has been fairly continuous when averaged over periods of tens of Myr, for more discussion see \citet{Kennicutt1998}).

We use these star formation rates to investigate the possibility of bursty star formation histories of these galaxies at moderate redshifts. We note that estimates based on Equations~\ref{eq:uvsfr} and \ref{eq:hasfr} assume a constant SFR for more than 100 Myr. However, for a robust comparison of H$\alpha$ and UV SFRs in this analysis, the absolute SFR scale is less important than a consistent set of calibrations. As H$\alpha$ and UV luminosities depend on the shape of the upper IMF for a given age, we expect the H$\alpha$ to UV ratio to vary no more than 30\% for typical IMF slopes \citep[e.g.][]{Glazebrook1999}.

\subsection{Structure and Morphologies}\label{sec:morfometryka}

We have used the \textsc{Morfometryka} application \citep{morfometryka} version 8.2 to measure the structure, non-parametric morphology, and Sersic profiles of the sources investigated here in the HST/F814W band \citep{Lotz2017}. We briefly describe how \textsc{Morfometryka} works below; please refer to \cite{morfometryka} for full details. 

\textsc{Morfometryka} takes as input the galaxy stamp image and the related PSF, then segments it and measures basic geometric parameters (e.g. centre, axis length, position angle). Next, it quantifies the radial light distribution I(R) from which the Petrosian Radius Rp \citep{Petrosian} and the half-light radius is estimated. For subsequent measurements, a Petrosian Region with the same geometric parameters as the galaxy and with a radius of 1.5 Rp is used. We also use this region to measure classic non-parametric morphology indicators, like the CAS system parameters \citep{Conselice2003a, Conselice2014}. Additionally, our method fits an 1D Sersic profile \citep{Sersic} for the quantified I(R) and uses the retrieved initial parameters to fit an 2D Sersic profile directly to the image of the central source. The structural and morphological measurements used in Section~ \ref{sec:structure_sizes} are summarised in Table~\ref{tab:results_morphology}, including errors estimated for the Sersic Index, Asymmetry, Concentration and Half-Light Radius.


\begin{table*}
\caption{Summary of the structure and morphology measurements of selected sample}
\label{tab:results_morphology}
\begin{tabular*}{\textwidth}{l@{\extracolsep{\fill}}cccccccccccr}
\hline \\ [-2mm]
ID  &                  R.A &                  DEC &  z &  SFR$_{\mathrm{H\alpha}}$  & M$_*$ &  n &   C &  A  &  R$_{50}$ & Label \\ [0.5mm]
    &                [deg] &                [deg] &    &  $\left[\log(\frac{\mathrm{M_{\odot}}}{yr})\right]$     & $\left[\log(\mathrm{\frac{M_*}{M_\odot}})\right]$ &    &                   &    &     & [kpc] &  \\ [0.5mm]
(1) &                  (2) &                (3)   & (4)& (5)    & (6)&  (7) & (8)& (9)& (10) & (11) \\ [1mm]
\hline \\ [-2mm]
443\hspace{15px}  &          40.0597647 &           -1.6489651 &    0.39 &  0.09$\pm$0.02  &  8.23 &   0.87 $\pm$ 0.03   &  3.16$\pm$0.07 & 0.08 $\pm$  0.01 & 2.75 $\pm$ 0.22 &         No UV \\
466  &          40.0561404 &           -1.6504371 &    0.25 &  0.20$\pm$0.03  & 10.36 &   5.09 $\pm$ 0.48   &  2.96$\pm$0.02 & 0.08 $\pm$  0.01 & 0.63 $\pm$ 0.14 &         No UV \\
492  &          40.0420305 &           -1.6473638 &    0.38 &  0.06$\pm$0.03  &  7.06 &   0.66 $\pm$ 0.16   &  2.69$\pm$0.30 & 0.08 $\pm$  0.05 & 1.14 $\pm$ 0.21 &         No UV \\
1060 &          39.9844793 &             -1.59586 &    0.36 &  0.02$\pm$0.01  &  7.37 &   1.05 $\pm$ 0.07   &  2.91$\pm$0.15 & 0.11 $\pm$  0.02 & 1.89 $\pm$ 0.20 &    Non-Bursty \\
1095 &         177.4125231 &           22.3825109 &    0.44 &  0.52$\pm$0.10  &  8.45 &   0.78 $\pm$ 0.08   &  2.46$\pm$0.08 & 0.10 $\pm$  0.07 & 1.98 $\pm$ 0.24 &    Non-Bursty \\
1247 &          39.9740886 &           -1.5945053 &    0.37 &  0.23$\pm$0.10  &  8.16 &   0.86 $\pm$ 0.03   &  2.56$\pm$0.07 & 0.15 $\pm$  0.01 & 2.16 $\pm$ 0.21 &    Non-Bursty \\
1612 &          39.9897845 &           -1.5903169 &    0.39 &  0.02$\pm$0.01  &  8.24 &   0.84 $\pm$ 0.02   &  3.05$\pm$0.13 & 0.12 $\pm$  0.03 & 2.85 $\pm$ 0.22 &         No UV \\
1706 &          39.9543907 &            -1.5893791 &   0.40 &  0.33$\pm$0.01  &  8.05 &   0.72 $\pm$ 0.03  &   2.50$\pm$0.15 & 0.07 $\pm$  0.02 & 2.13 $\pm$ 0.22 &         No UV \\
1796 &          39.9432867 &            -1.58851 &    0.38 & 0.08$\pm$0.02  &  8.72 &     1.86 $\pm$ 0.32  &   2.55$\pm$0.06 & 0.13 $\pm$  0.01 & 0.98 $\pm$ 0.21 &         No UV \\
2002 &          39.9821531 &            -1.5873661 &    0.39 &  0.05$\pm$0.01 &  8.09 &   0.81 $\pm$ 0.02  &   2.51$\pm$0.08 & 0.25 $\pm$  0.01 & 3.28 $\pm$ 0.22 &    Non-Bursty \\
2092 &          39.9575354 &            -1.5864846 &    0.42 &  0.86$\pm$0.39  &  9.37 &  0.87 $\pm$ 0.01   &  1.72$\pm$0.03 & 0.13 $\pm$  0.01 & 3.33 $\pm$ 0.23 &         No UV \\
2346 &          39.9581491 &           -1.5836976 &    0.38 &  0.02$\pm$0.01  &  8.57 &   1.04 $\pm$ 0.09   &  2.98$\pm$0.10 & 0.27 $\pm$  0.15 & 3.04 $\pm$ 0.21 &    Non-Bursty \\
2485 &          39.9732828 &           -1.5824717 &    0.44 &  0.42$\pm$0.04  &  8.50 &   0.80 $\pm$ 0.05   &  2.50$\pm$0.11 & 0.04 $\pm$  0.01 & 1.69 $\pm$ 0.24 &        Bursty \\
2699 &          39.9650007 &           -1.5815934 &    0.37 &  0.06$\pm$0.01  &  8.96 &   1.00 $\pm$ 0.04   &  2.96$\pm$0.06 & 0.03 $\pm$  0.01 & 1.72 $\pm$ 0.21 &    Non-Bursty \\
2921 &          39.9701364 &            -1.580751 &    0.36 &  0.27$\pm$0.10  &  9.03 &   1.19 $\pm$ 0.04   &  3.48$\pm$0.06 & 0.14 $\pm$  0.01 & 2.30 $\pm$ 0.20 &    Non-Bursty \\
2936 &          39.9732314 &            -1.580323 &    0.44 &  0.27$\pm$0.12  &  8.69 &   1.12 $\pm$ 0.04   &  2.69$\pm$0.10 & 0.04 $\pm$  0.01 & 2.58 $\pm$ 0.24 &    Non-Bursty \\
3265 &          40.0853888 &           -1.6243482 &    0.25 &  0.55$\pm$0.08 & 10.64 &    5.12 $\pm$ 0.49  &  2.67$\pm$0.01 &  0.08 $\pm$  0.01 & 0.69 $\pm$ 0.14 &         No UV \\
3328 &          39.9864872 &           -1.5763633 &    0.43 &  0.15$\pm$0.05  &  8.07 &   0.88 $\pm$ 0.05   &  2.59$\pm$0.15 & 0.07 $\pm$  0.03 & 2.03 $\pm$ 0.24 &    Non-Bursty \\
3367 &          39.9858426 &            -1.576675 &    0.39 &  0.17$\pm$0.08  &  8.66 &   1.25 $\pm$ 0.03   &  3.25$\pm$0.08 & 0.25 $\pm$  0.02 & 3.52 $\pm$ 0.22 &    Non-Bursty \\
3486 &          39.9711999 &            -1.5758353 &    0.43 &  0.25$\pm$0.13  &  8.92 &  0.97 $\pm$ 0.02   &  2.85$\pm$0.08 & 0.12 $\pm$  0.01 & 2.82 $\pm$ 0.24 &    Non-Bursty \\
3728 &          39.9727012 &           -1.5731656 &    0.39 &  0.02$\pm$0.01  &  8.42 &   0.76 $\pm$ 0.04   &  2.57$\pm$0.12 & 0.04 $\pm$  0.01 & 1.81 $\pm$ 0.22 &         No UV \\
3764 &          39.9825806 &           -1.5731697 &    0.43 &  0.45$\pm$0.35  &  8.30 &   0.73 $\pm$ 0.02   &  2.74$\pm$0.07 & 0.12 $\pm$  0.01 & 2.47 $\pm$ 0.24 &    Non-Bursty \\
3900 &          39.9676467 &           -1.5734048 &    0.45 &  1.01$\pm$0.53  &  9.25 &   1.01 $\pm$ 0.02   &  3.24$\pm$0.04 & 0.03 $\pm$  0.01 & 3.01 $\pm$ 0.25 &        Bursty \\
4033 &          39.9941189 &           -1.5710567 &    0.42 &  0.29$\pm$0.11  &  8.13 &   0.92 $\pm$ 0.05   &  2.70$\pm$0.12 & 0.08 $\pm$  0.02 & 2.17 $\pm$ 0.23 &    Non-Bursty \\
4080 &          177.4182231 &            22.3993438 &    0.41 &  1.69$\pm$0.75  &  9.56 & 1.82 $\pm$ 0.09   &  2.92$\pm$0.03 & 0.07 $\pm$  0.01 & 8.87 $\pm$ 0.23 &        Bursty \\
4371 &          39.9712901 &            -1.5709701 &    0.41 &  1.17$\pm$0.84  &  9.31 &  0.97 $\pm$ 0.03   &  2.74$\pm$0.05 & 0.02 $\pm$  0.01 & 1.89 $\pm$ 0.23 &        Bursty \\
4435 &          39.9811762 &            -1.568962 &    0.38 &  0.67$\pm$0.35  &  9.29 &   0.95 $\pm$ 0.04   &  3.29$\pm$0.06 & 0.06 $\pm$  0.01 & 1.69 $\pm$ 0.21 &         No UV \\
4476 &          39.9580378 &           -1.5674507 &    0.42 &  0.12$\pm$0.04  &  7.99 &   0.79 $\pm$ 0.04   &  2.44$\pm$0.16 & 0.05 $\pm$  0.04 & 2.49 $\pm$ 0.23 &    Non-Bursty \\
4652 &          39.9586683 &           -1.5663459 &    0.42 &  0.20$\pm$0.18  &  8.45 &   0.79 $\pm$ 0.02   &  2.62$\pm$0.10 & 0.04 $\pm$  0.01 & 2.46 $\pm$ 0.23 &         No UV \\
4661 &          39.9787517 &            -1.5666681 &    0.24 &  0.02$\pm$0.01  &  8.56 &  1.22 $\pm$ 0.02   &  3.45$\pm$0.07 & 0.32 $\pm$  0.01 & 2.23 $\pm$ 0.14 &    Non-Bursty \\
4702 &          39.9791276 &           -1.5654447 &    0.36 &  0.18$\pm$0.08  &  6.56 &   0.52 $\pm$ 0.03   &  2.43$\pm$0.15 & 0.37 $\pm$  0.04 & 2.25 $\pm$ 0.20 &    Non-Bursty \\
4742 &          177.4115124 &           22.4052383 &    0.39 &  0.05$\pm$0.03  &  8.40 &  4.19 $\pm$ 3.04   &  2.57$\pm$0.16 & 0.98 $\pm$  0.01 & 2.20 $\pm$ 0.22 &    Non-Bursty \\
5504 &         177.3925341 &           22.4123247 &    0.34 &  0.15$\pm$0.08  &  8.11 &   0.87 $\pm$ 0.04   &  3.29$\pm$0.06 & 0.38 $\pm$  0.09 & 2.92 $\pm$ 0.19 &    Non-Bursty \\
5615 &          39.9776652 &            -1.5567155 &    0.39 &  0.11$\pm$0.05  &  7.94 &  0.79 $\pm$ 0.03   &  3.51$\pm$0.10 & 0.23 $\pm$  0.01 & 3.65 $\pm$ 0.22 &    Non-Bursty \\
5867 &          39.9707607 &           -1.5541174 &    0.39 &  0.11$\pm$0.09  &  8.25 &   0.95 $\pm$ 0.07   &  2.67$\pm$0.15 & 0.16 $\pm$  0.01 & 1.54 $\pm$ 0.22 &         No UV \\
6061 &         177.3969102 &            22.4181006 &    0.36 &  0.16$\pm$0.12  &  9.28 &  1.02 $\pm$ 0.08   &  2.60$\pm$0.07 & 0.08 $\pm$  0.03 & 1.35 $\pm$ 0.20 &    Non-Bursty \\
6212 &          39.9729441 &            -1.5496746 &    0.36 &  0.41$\pm$0.09  &  9.00 &  0.90 $\pm$ 0.02   &  3.15$\pm$0.03 & 0.11 $\pm$  0.01 & 3.88 $\pm$ 0.20 &         No UV \\
6497 &          39.9760852 &            -1.5460288 &    0.36 &  1.18$\pm$0.39  &  9.96 &  1.06 $\pm$ 0.02   &  3.28$\pm$0.03 & 0.08 $\pm$  0.01 & 2.98 $\pm$ 0.20 &         No UV \\
[1mm] \hline \\ [-2mm]
\end{tabular*}

\begin{minipage}[c]{\textwidth}
\textbf{Table columns:} (1) HFF-DeepSpace catalogue ID; (2) Right ascension in degrees; (3) Declination in degrees; (4) Photometric redshift estimated using the EAZY software, including all available SHARDS and HFF-DeepSpace bands; (5) H$\alpha$ SFRs and associated error, calculated using the \citet{Kennicutt1998} conversion factor following corrections, as detailed in Section~\ref{sec:ha}; (6) Stellar mass in units of log solar masses, estimated with FAST utilising all available bands with typical errors of 0.2 dex; (7) Sersic index for a 2D profile (8) Concentration value; (9) Asymmetry value; (10) Half-Light Radius estimated from the radial light distribution I(R); (11) Label of star formation type.
\end{minipage}
\end{table*}


\section{Results}
\label{sec:res}

\subsection{Emission line candidates}

Using the methods outlined in Section~\ref{sec:cand}, we select emission line candidates based on the differential flux measurements using medium- and broad-band data. SHARDS fluxes are calibrated to match broad-band HST measurements, and are corrected for filter geometric effects (Equation~\ref{eq:cwl}) and broad-band continuum colour (Figure~\ref{fig:bbfix}). We then implement a two parameter selection criterion, based partially on agreement of line identification with photometric redshift, in order to minimise contamination from low redshift interlopers and the effects of random scatter. This selection process yields our candidate emission line galaxies from both cluster and their parallel fields (see Figure~\ref{fig:selection} and Section~\ref{sec:selection}).

One of the ways we identify correctly emission lines is by updating photometric redshift estimates through integrating all available SHARDS medium-band data. Using the SED fitting code EAZY \citep{Brammer2008} on a source-by-source basis, SHARDS filter response curves are shifted to the effective central wavelength of each object, allowing for robust photometric redshift estimates (see Figure~\ref{fig:comp}). In a similar manner, we update HFF-DeepSpace stellar population parameters using the FAST \citep{Kriek2009} code.

Our updated photometric redshift estimates allow for the identification of the emission lines responsible for the observed flux excess, as shown in Figure~\ref{fig:candidates}. Our sample is dominated by [\ion{O}{II}] and [\ion{O}{III}]+H$\beta$ emitters, as described earlier in the paper.  The photometric redshift criteria, along with the line emitter criteria ensures that our sample is a pure one with minimal contamination.   It may also appear that there is a continuous distribution of values, but this selection is done before we narrow down our targets, and many of these systems on the right hand side of Figure~7 are from rarer emission lines that are not often seen in shallower surveys.   Furthermore, we do find a concentration of objects at the expected common line emitter redshifts, and gaps at other redshifts, such as around $z \sim 1.1$.

While we find only 3 Lyman alpha candidates, one of which is a previously unknown $z\sim7$ object, which we will explore in future works.   As mentioned, this paper focuses on the H$\alpha$ emitters.    The reason for this is that the H$\alpha$ emitters are the closest systems to us and due to being the reddest emission line are affected less by dust and redshift effects. They therefore are easier to study, and give us some idea for how line emitters in the relatively nearby universe behave, which we can then compare with the higher redshift emitters.

In the next subsections we explore the properties of these H$\alpha$ emitters and give some idea of their origin.  This includes characterising their physical properties and comparing these to other known line emitters and field galaxies.  Specifically, we explore the star formation history of these objects and their structure, arguing that they are a population late time infall into cluster galaxies with induced star formation.


\subsection{Star formation main-sequence}

\begin{figure*}
	\includegraphics{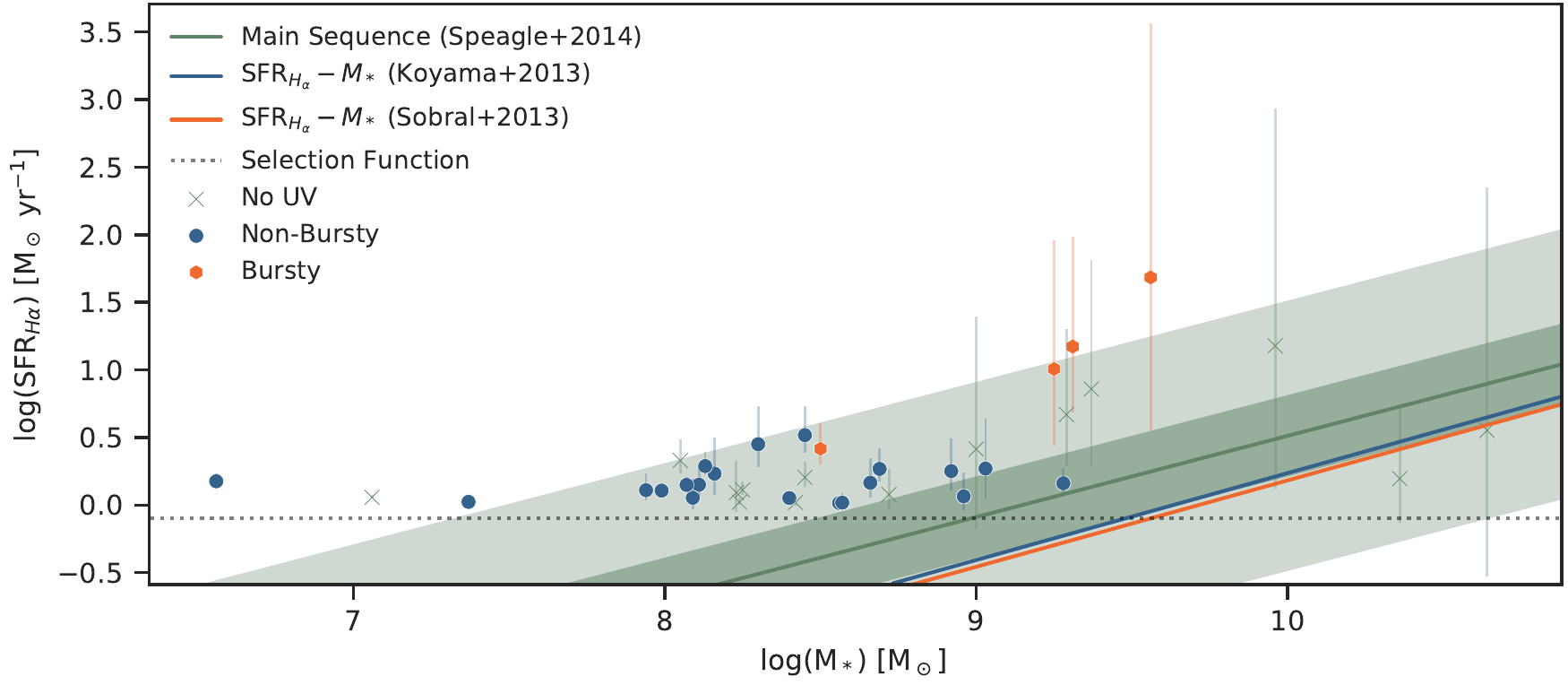}	
	\caption{Star formation vs. mass diagram for our z $\sim$ 0.4 line emitters with H$\alpha$ derived star formation rates. Black dotted line shows the selection function while the green line shows the main-sequence parameterization from \citet{Speagle2014} at z = 0.4. Dark, and light coloured bands have a width of a factor of 2 and 10 respectively. SFR$_{\rm{H\alpha}}$-M$_*$ relations of H$\alpha$ selected galaxies (at $z=0.4$) from HiZELS \citep{Sobral2013} and a rich cluster study \citep{Koyama2013} are shown as orange and blue solid lines respectively. Errors are extrapolated from uncertainties in luminosity and as such should be taken as lower bounds on the full error. See the text for a discussion of this trend under different assumptions on the dust extinction.}
	\label{fig:mainseq}
\end{figure*}

First, we investigate the star formation rates of our H$\alpha$ emitters and how these correlate with other properties. A crucial tool for understanding galaxy evolution is the star forming main sequence, or SFR-stellar mass relation. We present in Figure~\ref{fig:mainseq}, H$\alpha$ derived SFRs as a function of stellar mass. These H$\alpha$ SFRs are calculated using the \citet{Kennicutt1998} conversion factor following corrections as detailed in Section~\ref{sec:ha}, while stellar masses are measured using the SPS fitting code, FAST \citep{Kriek2009}. The green line in Figure~\ref{fig:mainseq} shows the z = 0.4 (redshift of our sample) main sequence parameterization of \citet{Speagle2014}, derived from rest-frame UV continuum based star formation rates.

In Figure~\ref{fig:mainseq} we also show the SFR$_{\rm{H\alpha}}$-M$_*$ relation for the $z=0.4$ H$\alpha$ emitter sample selected from the narrow-band High-Z Emission Line Survey \citep[HiZELS;][]{Sobral2013} in blue, and from the study of the rich cluster Cl 0939+4713 \citep{Koyama2013} in orange. 

As can be seen, we find a strong SFR$_{\rm{H\alpha}}$ enhancement of more than a factor of two for almost all galaxies in our sample over the \citet{Speagle2014} parameterization. This is indicative of a recent star burst in our galaxy sample's star formation history, the excess in SFR$_{\rm{H\alpha}}$ is further enhanced at low masses if one is to consider the \citet{Whitaker2014} parameterization, which finds a steep low mass slope and has been shown to be consistent with not only UV and IR star formation indicators, but also that of H$\alpha$. We note here however that care should be taken when making assumptions based on the \citet{Whitaker2014} parameterization as it is only constrained down to $log(M_*/M_{\odot}) = 8.4$.  Regardless, it is clear that these systems all have very large star formation rates for their mass and can be considered to be actively star forming galaxies, as expected given their identification as line emitters.  We later compare this H$\alpha$ star formation rate to the UV measured star formation rate to locate galaxies which exhibit `bursty' star formation histories. Note that as in \S 3.3.2 if we consider an equivalence between the extinction in stars and gas, rather than the 0.44 factor,  we would obtain a decrease in 0.26 dex in the log SFR axis.  This would however, not change the results discussed above. 

\subsection{Bursty star formation}
\label{sec:bursts}

From the previous sections it is clear that we have discovered a highly starforming population of moderate to low-mass galaxies at moderate redshifts.  The nature of these objects is however a mystery, and in the next sections we determine the properties of these systems, including their possible bursting nature, as well as their structure, environment and other properties.  

\subsubsection{Comparison of H$\alpha$ and UV derived SFRs}
\label{sec:sfr}

In this section we discuss the star formation rate history for our sample by comparing the H$\alpha$ to FUV fluxes.  Ideally, these two star formation rates should be the same if they measure the same aspects of star formation. However, for galaxies in which the star formation rate is more episodic or bursting, the measured H$\alpha$ SFR will often be higher than that measured from UV. The reason for this is that during a star formation episode, the star formation rate measured with H$\alpha$ will be higher during the initial burst than the UV measured star formation as detailed in Section~\ref{sec:intro}.  After some time, the H$\alpha$ measured star formation will become more similar to the UV measured one, especially for more constant star formation histories.

\begin{figure}
	\includegraphics{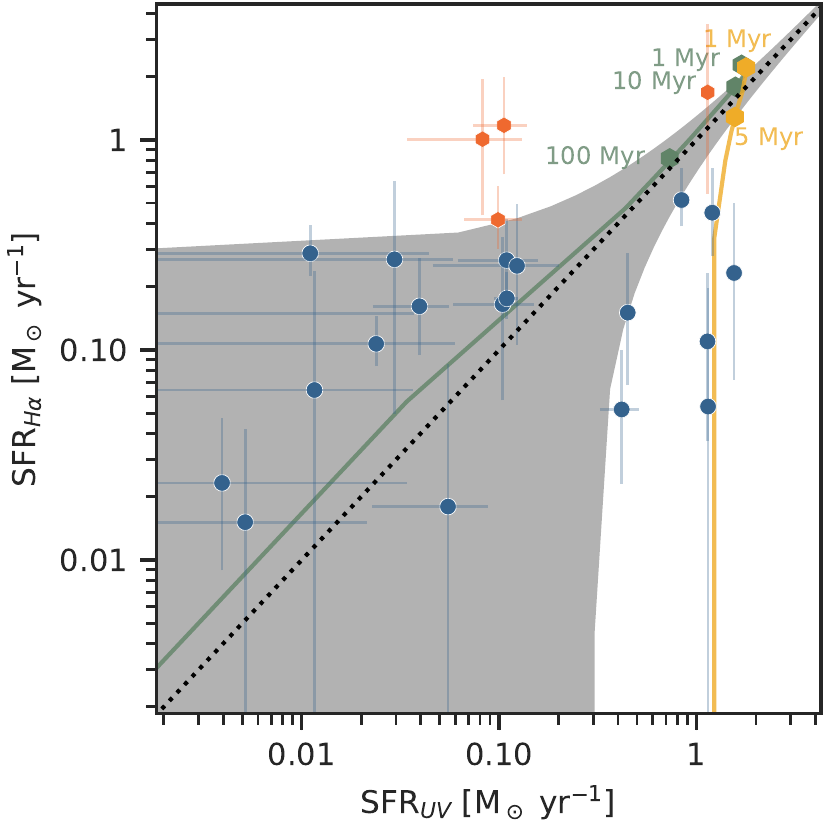}
	\caption{Comparison of the H$\alpha$ and UV SFRs for our sample of H$\alpha$ emitters. H$\alpha$ line fluxes are corrected for dust extinction and [\ion{N}{II}] line contamination, while the UV luminosities are dust corrected using stellar E(B-V) values derived from SED fitting. See the text for alternative parameterisations of these factors. Candidates from this study are marked by blue \& orange points and green crosses (as described in Fig~\ref{fig:mainseq}). Black dotted line shows a one-to-one ratio with a 0.3 dex margin, accounting for typical scatter in star formation activity due to mergers, gas flows or AGN feedback \citep{Tacchella2016}. We include two stellar population models from Starburst99 \citep{Leitherer1999} for reference. The green track shows a model stellar population with a constant star formation history of 1 M$_{\odot}$ yr$^{-1}$, with ages of 1 Myr, 10 Myr and 100 Myr marked by green points. While the yellow track shows a burst of star formation for a single stellar population with an initial mass of 10$^6$ M$_{\odot}$ and ages of 1 Myr and 5 Myr shown as yellow points.  Note that these models would need to be scale up in both directions to match the mass of individual galaxies in our sample by a factor of mass which is larger than 10$^{6}$ M$_{\odot}$. }
	\label{fig:SFRComp}
\end{figure}

In Figure~\ref{fig:SFRComp} we present a comparison of SFRs derived from H$\alpha$ and UV luminosities in order to provide a direct visualisation of possible bursty star formation activity. The H$\alpha$ line fluxes are initially computed via Equation~\ref{eq:fline} to account for a non-zero continuum before we apply corrections for dust extinction and [\ion{N}{II}] line contamination. UV luminosities are derived directly from reddening corrected rest-frame 1600\AA~emission, corresponding to the F275W band at $z \sim 0.4$. In order to obtain SFRs from H$\alpha$ and UV corrected luminosities we use the method described in Section~\ref{sec:sfrs}. To gain further insight into our sample and the possibility of bursty star formation histories, we include the evolutionary track of two example galaxies modelled with Starburst99\footnote{http://www.stsci.edu/science/starburst99/} \citep{Leitherer1999}. 

In yellow, we show the evolutionary track of an individual stellar population model of initial mass 10$^6$ M$_{\odot}$, undergoing a single burst of star formation, while model ages of 1 Myr, 5 Myr and 10 Myr are represented by the yellow points. While in green we show the evolutionary track of a model stellar population with a constant rate of star formation of 1 M$_{\odot}$ yr$^{-1}$ (green line), and indicate the SFR at ages of 1 Myr, 10 Myr and 100 Myr by the green points.

For simulated galaxies with a constant star formation history, H$\alpha$ derived SFRs usually match closely to values derived from the UV. In Figure~\ref{fig:SFRComp}, we show the one-to-one ratio of H$\alpha$ and UV SFRs, representing the equilibrium value of a constant star formation rate. The 0.3 dex margin accounts for variations around the main sequence due to changes in star formation activity, such as mergers, gas flows or AGN feedback \citep{Tacchella2016}. We find 65\% of our sample show an excess in H$\alpha$ SFR with respect to UV, while only 4 galaxies display significant enhancement of over 0.3 dex. Note that these numbers are still largely the same when we consider the dust extinction equivalence between stars and gas, as discussed in \S. 3.3.2. Only three more systems fall below the equivalence line in this case for this case of less dust extinction.  This is in indicative of a recent burst in star formation activity. We do however note that a number of systems exhibit SFR$_{\rm{H\alpha}}$/SFR$_{\rm{UV}}$ ratios of less than one, which can be attributed to galaxies in the post-burst phase, or consisting of a relatively young stellar population formed through a single burst of star formation. We discuss this further in Section~\ref{sec:ssfr} below.
   
There are however some important caveats in the direct comparison of SFRs, including the dust assumptions as already mentioned. Firstly, the conversion from luminosities to SFRs induce uncertainties due to unknown factors, such as; the initial mass function (IMF), stellar populations and metallicities \citep{Boselli2009} . Uncertainties in the IMF arise as a result of galaxies with a less abundant population of high-mass stars to ionize hydrogen, these objects will have similarly low ratios of H$\alpha$-to-FUV as those undergoing bursty star formation \citep{Lee2009}. For low-mass galaxies, these effects becomes particularly important, and luminosity to SFR conversions are expected to differ from the standard \citet{Kennicutt1998} prescriptions \citep{Ly2016}. Further sources of uncertainty arises from possibly more complicated dust parameterizations \citep{Kewley2002} and [\ion{N}{II}]/H$\alpha$ flux ratios. However, low-mass galaxies such as those in our sample are less affected by dust than higher redshift systems \citep{reddy2015}, while luminosity to SFR conversions and variations in [\ion{N}{II}]/H$\alpha$ line ratios are not sufficient to account for the excess in H$\alpha$ derived SFRs of our sample. 

\subsubsection{Specific star formation rates}
\label{sec:ssfr}

The relatively high SFR$_{\rm{H\alpha}}$/SFR$_{\rm{UV}}$ ratios present in our sample can be attributed to bursts of star formation, in which H$\alpha$ flux from short-lived O-stars is boosted with respect to that of the UV. However, it can also be explained by the assumption of  predominantly young galaxies formed in a single burst of star formation.  As can be seen in (Figure~\ref{fig:SFRComp}), models with a bursting star formation are able to match these values better than a constant star formation rate, except for the bursts which have a higher H$\alpha$ based star formation rate.

\begin{figure}
	\includegraphics{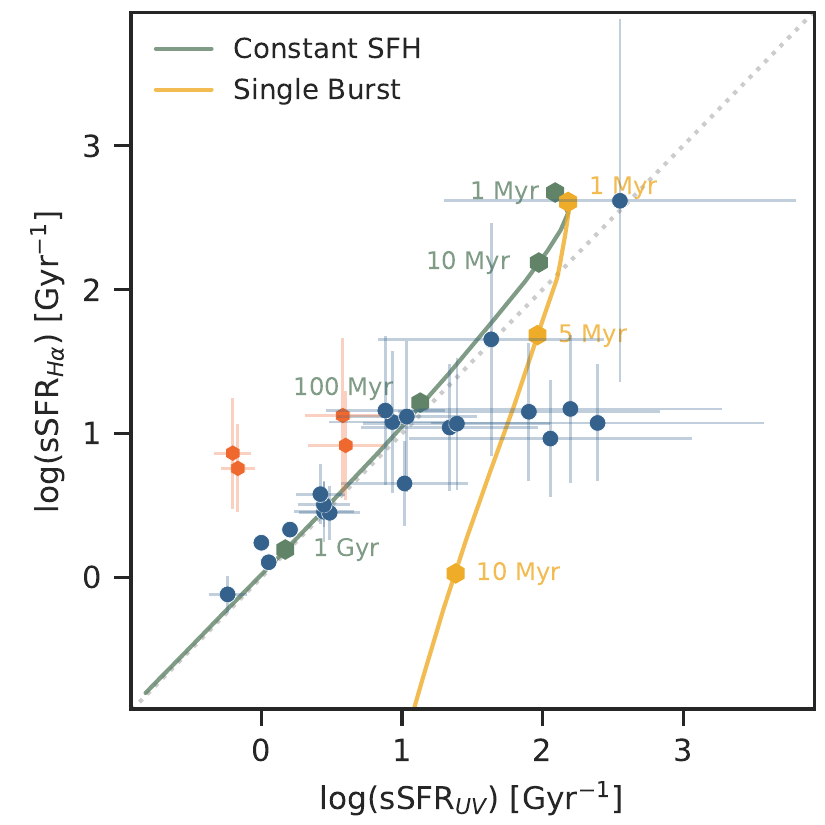}
	\caption{Specific star formation rates derived from H$\alpha$ and UV luminosities. These H$\alpha$ line fluxes are corrected for dust extinction and [\ion{N}{II}] line contamination (Section~\ref{sec:ha}) while UV luminosities are dust corrected using stellar E(B-V) values derived from SED fitting. Stellar masses used in the calculation of the sSFR are obtained through SPS modelling (see Section~\ref{sec:sps}). Candidates from this study are marked by blue \& orange points and green crosses (as described in Fig~\ref{fig:mainseq}). We include two stellar population models from Starburst99 \citep{Leitherer1999} for reference. The green track shows a model stellar population with a constant star formation history of 1 M$_{\odot}$ yr$^{-1}$, with ages of 1 Myr, 10 Myr, 100 Myr and 1 Gyr marked by green points. While the yellow track shows a single burst of star formation of a single stellar population and an initial mass of 10$^6$ M$_{\odot}$ and ages of 1 Myr, 5 Myr and 10 Myr shown as yellow points. }
	\label{fig:ssfr}
\end{figure}

In order to further investigate the origin of the observed enhancement in SFR$_{\rm{H\alpha}}$, we present the H$\alpha$ specific star formation rate (sSFR) as a function of UV sSFR in Figure~\ref{fig:ssfr}.  The sSFR is defined as the measured star formation rate divided by the stellar mass.  It gives some idea of the whether the star formation ongoing when a galaxy is observed is an important component of its mass formation.  Again, we show the evolutionary tracks of two Starburst99 models as discussed in Figure~\ref{fig:SFRComp}, and Section~\ref{sec:sfr}.


As shown by the evolutionary tracks in Figure~\ref{fig:ssfr}, the specific SFRs of a galaxy undergoing constant star formation (green line) eventually settles into equilibrium with little deviation. This is however not the case for a young galaxy formed under a single burst of star formation (yellow track), in which an enhancement in H$\alpha$ is expected for the first $\sim$5-10 Myr, but rapidly decreases with age as star formation is shut off. Thus, for a recent burst of star formation we would expect galaxies to be above the one-to-one relation of sSFR$_{\rm{H\alpha}}$ to sSFR$_{\rm{UV}}$. After some time these sources with ages $>$10 Myr after the burst would drop below this high excess. For our sample as a whole, we find no reason to favour young galaxy ages over bursty star formation histories as deviations from the one-to-one relation are typically within the margins expected due to perturbations such as gas flows or mergers. We note that $\sim$15\% of our sources exhibit a deficit in sSFR$_{\rm{UV}}$ over that of H$\alpha$ which could be attributed to young ages, but is more likely due to those galaxies being in the post-burst phase. These results would not change if we considered the alternative formalism for the dust extinction in gas compared to stars, as described in the previous section.

As can be seen in Figure~\ref{fig:ssfr}, our bursty sample, as defined by having a significantly larger SFR in H$\alpha$ than in the UV are not well fit by either the constant star formation or the single burst. However, if we consider the alternative equivalence between extinction in stars and gas, these points do approach the constant SFH models. This is however likely due to the fact that the bursts we see are occurring on top of an existing stellar mass.  From Figure~\ref{fig:ssfr}, we can see that the difference is about a factor of $\sim 30$, showing that the burst ongoing is about $\sim 3$\% of the galaxies mass, which would be consistent with an older system which is undergoing a new burst of star formation.

\subsection{Structure and Sizes}
\label{sec:structure_sizes}

We use the methodology explained in Section~\ref{sec:morfometryka} to measure the structures and sizes of our galaxies to better understand their origin in more detail. From the previous sections we know that these H$\alpha$ systems are undergoing star formation, with a small number undergoing bursts.   The sizes and structures of these systems may illuminate how the star formation within these systems triggered.

\begin{figure}
	\includegraphics[width=\columnwidth]{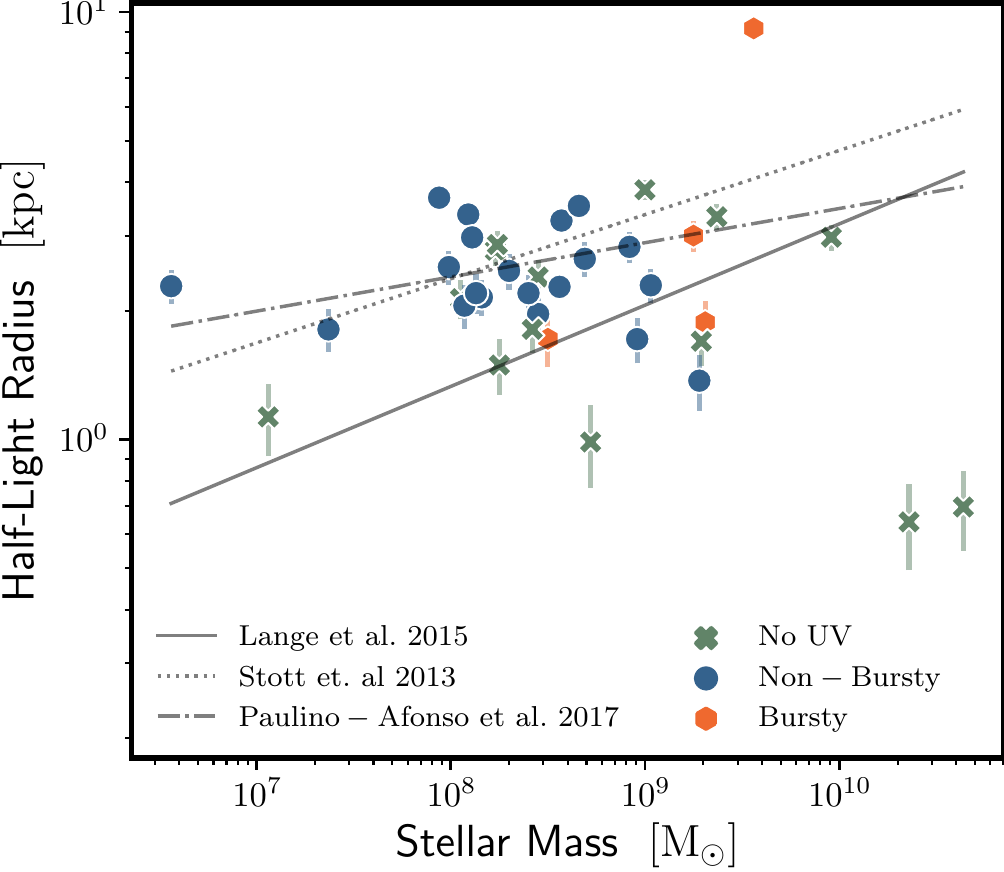}
	\caption{The size-mass relation for our sample of galaxies. Points are coloured according to the different methods in which our sample has been selected and measured. Solid, dotted and dash-dot lines show the size-mass relations from \citet{Lange2015}, \citet{Stott2013} and \citet{PaulinoAfonso2017}, respectively.}
	\label{fig:MRe}
\end{figure}

In Figure~\ref{fig:MRe} we show the distribution of the sizes vs. mass for our sample.  As can be seen, these systems are similar in size to galaxies of similar mass based on the size-mass relation for nearby galaxies \citep[from][]{Lange2015}. We also compare our objects to the size-mass relations derived from $z=0.4$ star-forming galaxies selected by H$\alpha$ emission. \citet{Stott2013} and \citet{PaulinoAfonso2017} both of whom contain samples of ($>400$) H$\alpha$ emitters from HiZELS \citep{Sobral2013}. From these size-mass relations (dotted and dashdot lines in Figure~\ref{fig:MRe}) we show that (with the exception of ID:4080) our galaxies are similar to typical H$\alpha$ emitters of corresponding mass and redshifts. 

\begin{figure}
	\includegraphics[width=\columnwidth]{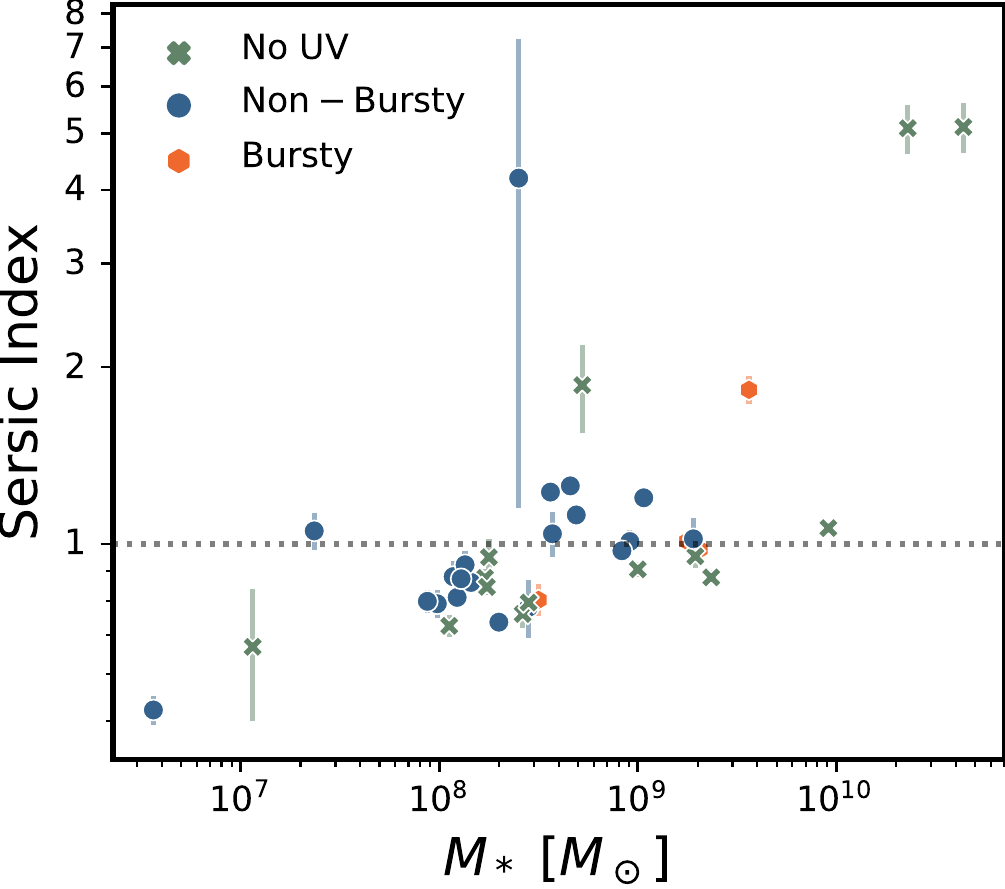}
	\caption{The relation between Sersic index, $n$ and the stellar mass.  Here we see the differences between our sample's structure with mass, showing that many of the bursty systems are found at the high-mass end of our relation, as well as have Sersic indices suggestive of disk galaxies. Sizes of the points are proportional to the $\tilde{\chi}^2$ of the Sersic profile fitting, with bigger points represent more uncertain fits.}
	\label{fig:nM}
\end{figure}

In terms of other aspects of the structure, we find that there are a range of Sersic indices for our systems, but that most of them have values of around $n=1$, suggesting that these systems are more disk-like, rather than spheroidal like.  This can also be gleamed by investigating the visual morphologies of these systems shown in Figure~\ref{fig:nM} where most of them appear to be disk-like objects rather than ellipticals.

\begin{figure}
	\includegraphics[width=\columnwidth]{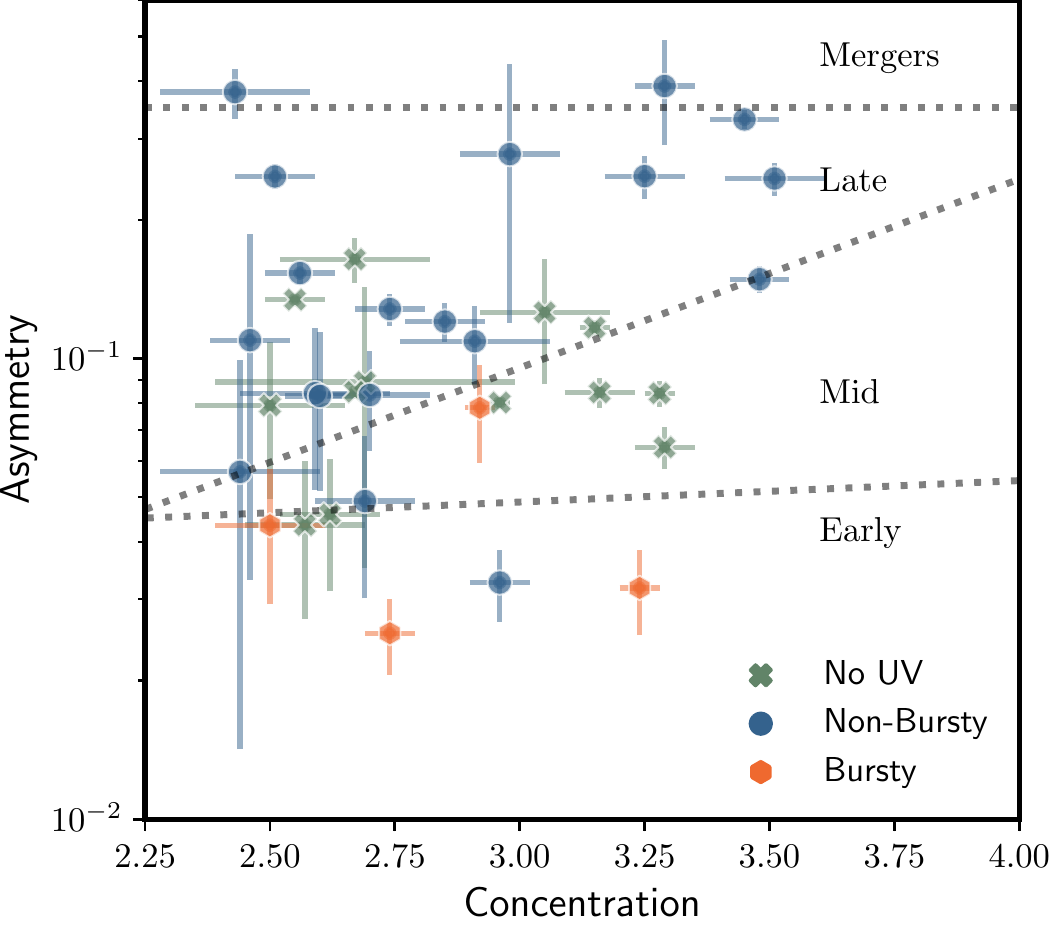}
	\caption{Diagram showing the concentration vs. asymmetry diagram for our sample. The points are defined in the same way as in Figure~\ref{fig:MRe}.  The lines denoting the differences between various $z = 0$ systems is shown from \citet{Bershady2000}. As can be seen very few to none of our systems have a morphology consistent with undergoing major mergers.}
	\label{fig:AC}
\end{figure}

\begin{figure*}
    \centering
    \includegraphics[width=\textwidth]{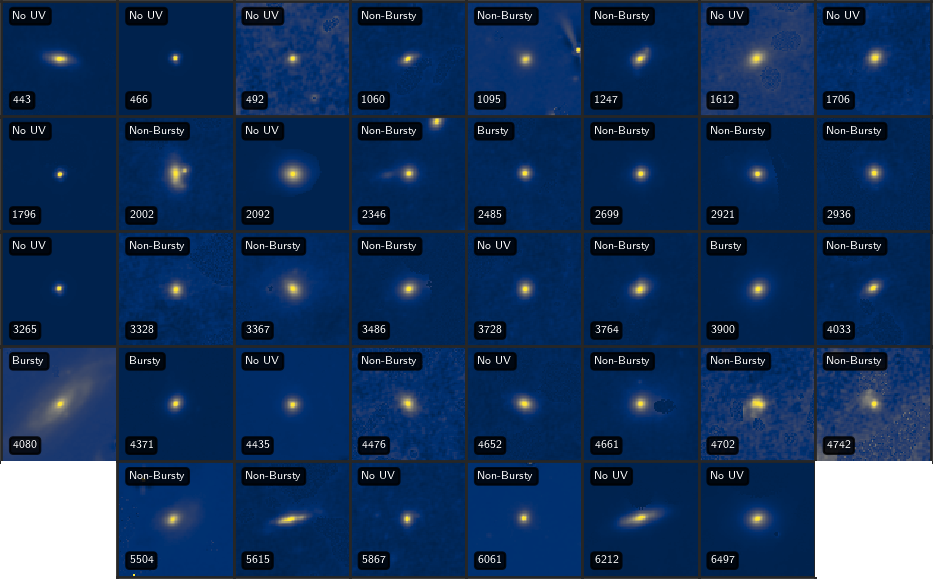}
    \caption{Mosaic of the 38 sources in the HST/F814W band \citep{Lotz2017} with structural and morphology measurements summarised in Table~\ref{tab:results_morphology}, the unique identifier of each object is shown in the bottom left corner for reference. A label using the same designation of the plots with No UV/Non-Bursty/Bursty is shown in the top left. Images presented here are the input stamps used with \textsc{Morfometryka}, external sources were extracted and replaced with a noise matching the stamp noise level.}
    \label{fig:mosaic}
\end{figure*}

Furthermore, the bursting galaxies, quite interestingly,  do not show any signs of being mergers.  They are not found in the merger space of the C-A diagram (Figure~\ref{fig:AC}), and do not appear to have a merger morphology when examined by eye (see Figure~\ref{fig:mosaic}). This implies that these systems do not have their star formation triggered directly by merging activity.  In fact, the bursts themselves are amongst the lowest asymmetry objects in the early-type galaxy region.   This is a strong sign that other mechanisms, besides merging, are responsible for producing the star formation within these systems.

\section{Discussion}
\label{sec:discussion}

We find a significant number (42) low-mass star forming galaxies selected by H$\alpha$ emission in narrow band filters though deep imaging of small pointed fields. Our analysis shows that all of these systems are above the star formation main sequence, and thus have an enhancement of star formation. This is likely partially due to our procedure for finding such systems, as we are more likely to find higher star formation rates in H$\alpha$ given the requirement that a significant detection exists within our medium-band imaging. Regardless, this is an interesting, highly starforming population of galaxies. We discover that about half of our systems can be defined as `bursts' which have a significantly higher star formation rate in the H$\alpha$ compared with the UV measured star formation rate.  We show that these systems can be due to very young starbursts within star formation events which occurred in the the past few Myr.  Therefore, we have discovered that some of these galaxies have their star formation triggered by a recent galaxy formation event.

One issue that we have not discussed in any detail in this paper, thus far, is that many of our objects, but not all, are within the cluster environment.  As Figure~\ref{fig:locations} shows, the distribution of our galaxies is such that many of them are within the A0370 cluster.  This cluster is at the redshift of our sources and it is likely that these galaxies are within the cluster itself.  The fact that we see more of these systems within a cluster than without might be the result of just having an overdensity in clusters compared to field galaxies, with a steeper slope mass function \citep[e.g.][]{Penny2008}.  However, it is worth considering that the cluster environment produces or influences the star formation within these galaxies, in particular for bursts.    As can be seen from the images in Figure~\ref{fig:locations}, many of our bursts are near larger galaxies.  This may be a sign that interactions, even perhaps high-speed ones such as galaxy harassment \citep[e.g.][]{Moore1998} are producing this effect.  Similar results are found by \citet{Stroe2017} in which higher densities of H$\alpha$ emitters are found in merging clusters over more relaxed systems. Further, \citet{Dressler1983} found signatures of increased star formation in the spectra of cluster galaxies, thought to be induced by ram pressure and later confirmed by the simulations of \citet{Bekki2003}. More recent simulations propose large-scale, low pressure shocks induced by merger processes could trigger star formation in cluster galaxies \citep{Roediger2014}.

When we examine the morphologies of these galaxies (Figure~\ref{fig:nM}) we find that most of them appear to to disk-like, which is consistent with their low stellar masses.  One of the burst galaxies appears to be a spiral (ID:4080), which may have recently fallen into the cluster, but the other three do not seem to have any detailed substructure. However, as we show and discuss in Section~\ref{sec:structure_sizes}, these systems are of typical size for their mass.     We conclude that therefore these systems are likely in some type of evolutionary stage whereby infalling field galaxies are undergoing rapid star formation, likely due to the cluster environment triggering this.  On the other hand they could be galaxies stripped of mass by the cluster environment, but it is unlikely that this process would be efficient enough to produce so many systems.    Future observations of these galaxies, through e.g., spectroscopy, should help us to better understand the origins of these systems which appear to be a new class of low mass galaxy.

\section{Conclusions}
\label{sec:conc}

In this paper we investigate the selection of emission line galaxies from two of the Hubble Frontier Fields clusters using SHARDS medium-band imaging.  Our Major findings are:

I. With the full suite of SHARDS observations now available (for A0370), we carry out an investigation to  discover and study emission line selected galaxies over a wide redshift range. While lacking the wide area coverage studies such as this typically use, deep imaging through the full set of 25 contiguous medium-band filters combined with the strong lensing potential of the Frontier Field clusters provides an essential tool for the identification and study of line emitters at a variety of redshifts. Particularly those at high redshifts which would typically be too faint to be detected in field observations or observations carried out with smaller ground based telescopes. 

II. We identify 1,098 candidate emission line galaxies based on a modified version of a well established two parameter selection criterion. SHARDS photometry is calibrated to match HST images and used to update photometric redshift estimates and stellar population parameters. Using a strict criteria, we are able to match a fraction of these to well known emission lines, such as H$\alpha$, the focus of this paper. Many of the other sources are likely from more obscure emission lines.

III. We discover 38 predominantly mid to low-mass H$\alpha$ emitters, which is the focus of the latter parts of this paper.  After correcting for dust extinction and [\ion{N}{II}] line contamination we derive H$\alpha$-based star formation rates. Overall, 27 of these candidates have corresponding HST UV data which enables us to investigate potential bursts of star formation in our galaxies recent histories by comparing the two indicators.  Most of our H$\alpha$ emitters are undergoing star formation at high enough values such that these galaxies fall above the main sequence relation between star formation and stellar mass.

IV. We investigate  bursty star formation histories for these star forming galaxies using different star formation indicators that trace different time scales. We find a significant enhancement ($>0.3$ dex) of SFR$_{\rm{H\alpha}}$ over that of SFR$_{\rm{UV}}$ for 4 of our candidates. This fact, implies a recent burst in star formation within these systems.  These systems are likely within the foreground Frontier Fields clusters, and thus their bursty star formation is probably induced by cluster processes. This result is robust to different considerations of the dust extinction we use to correct the H$\alpha$ based star formation rates.

V. Enhancements in the H$\alpha$ SFR are typically attributed to bursty star formation, but can also result from a young population of galaxies. In order to differentiate we compare the H$\alpha$ and UV sSFR to the evolutionary tracks of two model galaxies. We find no reason to favour a young first generation population for the majority of our sample, but note that up to 10\% of our galaxies show a tentative indication of young ages over bursty star formation histories.  

We conclude that SHARDS imaging, combined with deep Hubble Space Telescope data set can successfully identify emission line galaxies over a wide range of redshifts. Future papers will investigate the higher redshift emitters in a similar way.   These methods, coupled with the strong lensing potential of the Frontier Fields clusters enables the identification of a population of low mass, faint galaxies likely undergoing a burst in star formation.  These objects are fairly unique and should be followed up as they are likely tracing important processes in the formation of galaxies within dense environments.  

\section*{Acknowledgements}

This work was supported by the Science and Technology Facilities Council in the form of a studentship to AG. LF acknowledges funding from Coordena\c{c}\~{a}o de Aperfei\c{c}oamento de Pessoal de N\'{i}vel Superior - Brasil (CAPES) - Finance Code 001. DRG thanks CONACyT for the research grant CB-A1-S-22784. This paper is (partly) based on SHARDS-FF data. SHARDS-FF is currently funded by Spanish Government grant PGC2018-093499-BI00. Based on observations made with the Gran Telescopio Canarias (GTC), installed at the Spanish Observatorio del Roque de los Muchachos of the Instituto de Astrof\'{i}sica de Canarias, in the island of La Palma.  This work is also based partially on data and catalog products from HFF-DeepSpace, funded by the National Science Foundation and Space Telescope Science Institute (operated by the Association of Universities for Research in Astronomy, Inc., under NASA contract NAS5-26555).

\section*{Data availability}
The data underlying this article will be shared on reasonable request to the corresponding author.




\bibliographystyle{mnras}
\bibliography{reference}

\begin{thebibliography}{}
\makeatletter
\relax
\def\mn@urlcharsother{\let\do\@makeother \do\$\do\&\do\#\do\^\do\_\do\%\do\~}
\def\mn@doi{\begingroup\mn@urlcharsother \@ifnextchar [ {\mn@doi@}
  {\mn@doi@[]}}
\def\mn@doi@[#1]#2{\def\@tempa{#1}\ifx\@tempa\@empty \href
  {http://dx.doi.org/#2} {doi:#2}\else \href {http://dx.doi.org/#2} {#1}\fi
  \endgroup}
\def\mn@eprint#1#2{\mn@eprint@#1:#2::\@nil}
\def\mn@eprint@arXiv#1{\href {http://arxiv.org/abs/#1} {{\tt arXiv:#1}}}
\def\mn@eprint@dblp#1{\href {http://dblp.uni-trier.de/rec/bibtex/#1.xml}
  {dblp:#1}}
\def\mn@eprint@#1:#2:#3:#4\@nil{\def\@tempa {#1}\def\@tempb {#2}\def\@tempc
  {#3}\ifx \@tempc \@empty \let \@tempc \@tempb \let \@tempb \@tempa \fi \ifx
  \@tempb \@empty \def\@tempb {arXiv}\fi \@ifundefined
  {mn@eprint@\@tempb}{\@tempb:\@tempc}{\expandafter \expandafter \csname
  mn@eprint@\@tempb\endcsname \expandafter{\@tempc}}}

\bibitem[\protect\citeauthoryear{{Arrabal Haro} et~al.,}{{Arrabal Haro}
  et~al.}{2018}]{arrabal2018}
{Arrabal Haro} P.,  et~al., 2018, \mn@doi [\mnras] {10.1093/mnras/sty1106},
  \href {https://ui.adsabs.harvard.edu/abs/2018MNRAS.478.3740A} {478, 3740}

\bibitem[\protect\citeauthoryear{{Banerji} et~al.,}{{Banerji}
  et~al.}{2013}]{Banerji2013}
{Banerji} M.,  et~al., 2013, \mn@doi [\mnras] {10.1093/mnras/stt320}, \href
  {https://ui.adsabs.harvard.edu/abs/2013MNRAS.431.2209B} {431, 2209}

\bibitem[\protect\citeauthoryear{{Beckwith} et~al.,}{{Beckwith}
  et~al.}{2006}]{Beckwith2006}
{Beckwith} S. V.~W.,  et~al., 2006, \mn@doi [\aj] {10.1086/507302}, \href
  {https://ui.adsabs.harvard.edu/abs/2006AJ....132.1729B} {132, 1729}

\bibitem[\protect\citeauthoryear{{Bekki} \& {Couch}}{{Bekki} \&
  {Couch}}{2003}]{Bekki2003}
{Bekki} K.,  {Couch} W.~J.,  2003, \mn@doi [\apjl] {10.1086/379054}, \href
  {https://ui.adsabs.harvard.edu/abs/2003ApJ...596L..13B} {596, L13}

\bibitem[\protect\citeauthoryear{{Bershady}, {Jangren}  \&
  {Conselice}}{{Bershady} et~al.}{2000}]{Bershady2000}
{Bershady} M.~A.,  {Jangren} A.,   {Conselice} C.~J.,  2000, \mn@doi [\aj]
  {10.1086/301386}, \href
  {https://ui.adsabs.harvard.edu/abs/2000AJ....119.2645B} {119, 2645}

\bibitem[\protect\citeauthoryear{{Bertin} \& {Arnouts}}{{Bertin} \&
  {Arnouts}}{1996}]{Bertin1996}
{Bertin} E.,  {Arnouts} S.,  1996, \mn@doi [\aaps] {10.1051/aas:1996164}, \href
  {http://adsabs.harvard.edu/abs/1996A%26AS..117..393B} {117, 393}

\bibitem[\protect\citeauthoryear{{Boselli}, {Boissier}, {Cortese}, {Buat},
  {Hughes}  \& {Gavazzi}}{{Boselli} et~al.}{2009}]{Boselli2009}
{Boselli} A.,  {Boissier} S.,  {Cortese} L.,  {Buat} V.,  {Hughes} T.~M.,
  {Gavazzi} G.,  2009, \mn@doi [\apj] {10.1088/0004-637X/706/2/1527}, \href
  {https://ui.adsabs.harvard.edu/abs/2009ApJ...706.1527B} {706, 1527}

\bibitem[\protect\citeauthoryear{{Brammer}, {van Dokkum}  \& {Coppi}}{{Brammer}
  et~al.}{2008}]{Brammer2008}
{Brammer} G.~B.,  {van Dokkum} P.~G.,   {Coppi} P.,  2008, \mn@doi [\apj]
  {10.1086/591786}, \href {http://adsabs.harvard.edu/abs/2008ApJ...686.1503B}
  {686, 1503}

\bibitem[\protect\citeauthoryear{{Bruzual} \& {Charlot}}{{Bruzual} \&
  {Charlot}}{2003}]{Bruzual2003}
{Bruzual} G.,  {Charlot} S.,  2003, \mn@doi [\mnras]
  {10.1046/j.1365-8711.2003.06897.x}, \href
  {https://ui.adsabs.harvard.edu/abs/2003MNRAS.344.1000B} {344, 1000}

\bibitem[\protect\citeauthoryear{{Bunker}, {Warren}, {Hewett}  \&
  {Clements}}{{Bunker} et~al.}{1995}]{Bunker1995}
{Bunker} A.~J.,  {Warren} S.~J.,  {Hewett} P.~C.,   {Clements} D.~L.,  1995,
  \mn@doi [\mnras] {10.1093/mnras/273.2.513}, \href
  {http://adsabs.harvard.edu/abs/1995MNRAS.273..513B} {273, 513}

\bibitem[\protect\citeauthoryear{{Calzetti}, {Armus}, {Bohlin}, {Kinney},
  {Koornneef}  \& {Storchi-Bergmann}}{{Calzetti} et~al.}{2000}]{Calzetti2000}
{Calzetti} D.,  {Armus} L.,  {Bohlin} R.~C.,  {Kinney} A.~L.,  {Koornneef} J.,
   {Storchi-Bergmann} T.,  2000, \mn@doi [\apj] {10.1086/308692}, \href
  {https://ui.adsabs.harvard.edu/abs/2000ApJ...533..682C} {533, 682}

\bibitem[\protect\citeauthoryear{{Ceverino}, {S{\'a}nchez Almeida}, {Mu{\~n}oz
  Tu{\~n}{\'o}n}, {Dekel}, {Elmegreen}, {Elmegreen}  \& {Primack}}{{Ceverino}
  et~al.}{2016a}]{Ceverino2016a}
{Ceverino} D.,  {S{\'a}nchez Almeida} J.,  {Mu{\~n}oz Tu{\~n}{\'o}n} C.,
  {Dekel} A.,  {Elmegreen} B.~G.,  {Elmegreen} D.~M.,   {Primack} J.,  2016a,
  \mn@doi [\mnras] {10.1093/mnras/stw064}, \href
  {https://ui.adsabs.harvard.edu/abs/2016MNRAS.457.2605C} {457, 2605}

\bibitem[\protect\citeauthoryear{{Ceverino}, {Arribas}, {Colina},
  {Rodr{\'\i}guez Del Pino}, {Dekel}  \& {Primack}}{{Ceverino}
  et~al.}{2016b}]{Ceverino2016b}
{Ceverino} D.,  {Arribas} S.,  {Colina} L.,  {Rodr{\'\i}guez Del Pino} B.,
  {Dekel} A.,   {Primack} J.,  2016b, \mn@doi [\mnras] {10.1093/mnras/stw1195},
  \href {https://ui.adsabs.harvard.edu/abs/2016MNRAS.460.2731C} {460, 2731}

\bibitem[\protect\citeauthoryear{{Ceverino}, {Klessen}  \& {Glover}}{{Ceverino}
  et~al.}{2018}]{Ceverino2018}
{Ceverino} D.,  {Klessen} R.~S.,   {Glover} S. C.~O.,  2018, \mn@doi [\mnras]
  {10.1093/mnras/sty2124}, \href
  {https://ui.adsabs.harvard.edu/abs/2018MNRAS.480.4842C} {480, 4842}

\bibitem[\protect\citeauthoryear{{Chabrier}}{{Chabrier}}{2003}]{Chabrier2003}
{Chabrier} G.,  2003, \mn@doi [\pasp] {10.1086/376392}, \href
  {https://ui.adsabs.harvard.edu/abs/2003PASP..115..763C} {115, 763}

\bibitem[\protect\citeauthoryear{{Conroy} \& {Gunn}}{{Conroy} \&
  {Gunn}}{2010}]{Conroy2010}
{Conroy} C.,  {Gunn} J.~E.,  2010, \mn@doi [\apj]
  {10.1088/0004-637X/712/2/833}, \href
  {https://ui.adsabs.harvard.edu/abs/2010ApJ...712..833C} {712, 833}

\bibitem[\protect\citeauthoryear{{Conroy}, {Gunn}  \& {White}}{{Conroy}
  et~al.}{2009}]{Conroy2009}
{Conroy} C.,  {Gunn} J.~E.,   {White} M.,  2009, \mn@doi [\apj]
  {10.1088/0004-637X/699/1/486}, \href
  {https://ui.adsabs.harvard.edu/abs/2009ApJ...699..486C} {699, 486}

\bibitem[\protect\citeauthoryear{{Conselice}}{{Conselice}}{2003}]{Conselice2003a}
{Conselice} C.~J.,  2003, \mn@doi [\apjs] {10.1086/375001}, \href
  {https://ui.adsabs.harvard.edu/abs/2003ApJS..147....1C} {147, 1}

\bibitem[\protect\citeauthoryear{Conselice}{Conselice}{2014}]{Conselice2014}
Conselice C.~J.,  2014, \mn@doi [Annual Review of Astronomy and Astrophysics]
  {10.1146/annurev-astro-081913-040037}, 52, 291

\bibitem[\protect\citeauthoryear{{Conselice}, {Bershady}, {Dickinson}  \&
  {Papovich}}{{Conselice} et~al.}{2003}]{Conselice2003}
{Conselice} C.~J.,  {Bershady} M.~A.,  {Dickinson} M.,   {Papovich} C.,  2003,
  \mn@doi [\aj] {10.1086/377318}, \href
  {https://ui.adsabs.harvard.edu/abs/2003AJ....126.1183C} {126, 1183}

\bibitem[\protect\citeauthoryear{{Conselice}, {Wilkinson}, {Duncan}  \&
  {Mortlock}}{{Conselice} et~al.}{2016}]{Conselice2016}
{Conselice} C.~J.,  {Wilkinson} A.,  {Duncan} K.,   {Mortlock} A.,  2016,
  \mn@doi [\apj] {10.3847/0004-637X/830/2/83}, \href
  {https://ui.adsabs.harvard.edu/abs/2016ApJ...830...83C} {830, 83}

\bibitem[\protect\citeauthoryear{{Dahlen} et~al.,}{{Dahlen}
  et~al.}{2013}]{Dahlen2013}
{Dahlen} T.,  et~al., 2013, \mn@doi [\apj] {10.1088/0004-637X/775/2/93}, \href
  {https://ui.adsabs.harvard.edu/abs/2013ApJ...775...93D} {775, 93}

\bibitem[\protect\citeauthoryear{{Dalton} et~al.,}{{Dalton}
  et~al.}{2020}]{weave2020}
{Dalton} G.,  et~al., 2020, in Society of Photo-Optical Instrumentation
  Engineers (SPIE) Conference Series. p. 1144714, \mn@doi{10.1117/12.2561067}

\bibitem[\protect\citeauthoryear{{Dom{\'\i}nguez}, {Siana}, {Brooks},
  {Christensen}, {Bruzual}, {Stark}  \& {Alavi}}{{Dom{\'\i}nguez}
  et~al.}{2015}]{Dominguez2015}
{Dom{\'\i}nguez} A.,  {Siana} B.,  {Brooks} A.~M.,  {Christensen} C.~R.,
  {Bruzual} G.,  {Stark} D.~P.,   {Alavi} A.,  2015, \mn@doi [\mnras]
  {10.1093/mnras/stv1001}, \href
  {https://ui.adsabs.harvard.edu/abs/2015MNRAS.451..839D} {451, 839}

\bibitem[\protect\citeauthoryear{{Dressler} \& {Gunn}}{{Dressler} \&
  {Gunn}}{1983}]{Dressler1983}
{Dressler} A.,  {Gunn} J.~E.,  1983, \mn@doi [\apj] {10.1086/161093}, \href
  {https://ui.adsabs.harvard.edu/abs/1983ApJ...270....7D} {270, 7}

\bibitem[\protect\citeauthoryear{{Duncan} \& {Conselice}}{{Duncan} \&
  {Conselice}}{2015}]{Duncan2015}
{Duncan} K.,  {Conselice} C.~J.,  2015, \mn@doi [\mnras]
  {10.1093/mnras/stv1049}, \href
  {https://ui.adsabs.harvard.edu/abs/2015MNRAS.451.2030D} {451, 2030}

\bibitem[\protect\citeauthoryear{{Duncan} et~al.,}{{Duncan}
  et~al.}{2019}]{Duncan2019}
{Duncan} K.,  et~al., 2019, \mn@doi [\apj] {10.3847/1538-4357/ab148a}, \href
  {https://ui.adsabs.harvard.edu/abs/2019ApJ...876..110D} {876, 110}

\bibitem[\protect\citeauthoryear{{Emami}, {Siana}, {Weisz}, {Johnson}, {Ma}  \&
  {El-Badry}}{{Emami} et~al.}{2019}]{Emami2019}
{Emami} N.,  {Siana} B.,  {Weisz} D.~R.,  {Johnson} B.~D.,  {Ma} X.,
  {El-Badry} K.,  2019, \mn@doi [\apj] {10.3847/1538-4357/ab211a}, \href
  {https://ui.adsabs.harvard.edu/abs/2019ApJ...881...71E} {881, 71}

\bibitem[\protect\citeauthoryear{{Ferrari}, {de Carvalho}  \&
  {Trevisan}}{{Ferrari} et~al.}{2015}]{morfometryka}
{Ferrari} F.,  {de Carvalho} R.~R.,   {Trevisan} M.,  2015, \mn@doi [\apj]
  {10.1088/0004-637X/814/1/55}, \href
  {https://ui.adsabs.harvard.edu/abs/2015ApJ...814...55F} {814, 55}

\bibitem[\protect\citeauthoryear{{Garn} \& {Best}}{{Garn} \&
  {Best}}{2010}]{Garn2010b}
{Garn} T.,  {Best} P.~N.,  2010, \mn@doi [\mnras]
  {10.1111/j.1365-2966.2010.17321.x}, \href
  {https://ui.adsabs.harvard.edu/abs/2010MNRAS.409..421G} {409, 421}

\bibitem[\protect\citeauthoryear{{Geach}, {Smail}, {Best}, {Kurk}, {Casali},
  {Ivison}  \& {Coppin}}{{Geach} et~al.}{2008}]{Geach2008}
{Geach} J.~E.,  {Smail} I.,  {Best} P.~N.,  {Kurk} J.,  {Casali} M.,  {Ivison}
  R.~J.,   {Coppin} K.,  2008, \mn@doi [\mnras]
  {10.1111/j.1365-2966.2008.13481.x}, \href
  {https://ui.adsabs.harvard.edu/abs/2008MNRAS.388.1473G} {388, 1473}

\bibitem[\protect\citeauthoryear{{Glazebrook}, {Blake}, {Economou}, {Lilly}  \&
  {Colless}}{{Glazebrook} et~al.}{1999}]{Glazebrook1999}
{Glazebrook} K.,  {Blake} C.,  {Economou} F.,  {Lilly} S.,   {Colless} M.,
  1999, \mn@doi [\mnras] {10.1046/j.1365-8711.1999.02576.x}, \href
  {https://ui.adsabs.harvard.edu/abs/1999MNRAS.306..843G} {306, 843}

\bibitem[\protect\citeauthoryear{{Grogin} et~al.,}{{Grogin}
  et~al.}{2011}]{Grogin2011}
{Grogin} N.~A.,  et~al., 2011, \mn@doi [\apjs] {10.1088/0067-0049/197/2/35},
  \href {https://ui.adsabs.harvard.edu/abs/2011ApJS..197...35G} {197, 35}

\bibitem[\protect\citeauthoryear{{Gronke}, {Dijkstra}, {McCourt}  \&
  {Oh}}{{Gronke} et~al.}{2016}]{Gronke2016}
{Gronke} M.,  {Dijkstra} M.,  {McCourt} M.,   {Oh} S.~P.,  2016, \mn@doi
  [\apjl] {10.3847/2041-8213/833/2/L26}, \href
  {https://ui.adsabs.harvard.edu/abs/2016ApJ...833L..26G} {833, L26}

\bibitem[\protect\citeauthoryear{{Guhathakurta}, {Tyson}  \&
  {Majewski}}{{Guhathakurta} et~al.}{1990}]{Guhathakurta1990}
{Guhathakurta} P.,  {Tyson} J.~A.,   {Majewski} S.~R.,  1990, \mn@doi [\apjl]
  {10.1086/185754}, \href
  {https://ui.adsabs.harvard.edu/abs/1990ApJ...357L...9G} {357, L9}

\bibitem[\protect\citeauthoryear{{Ibar} et~al.,}{{Ibar}
  et~al.}{2013}]{Ibar2013}
{Ibar} E.,  et~al., 2013, \mn@doi [\mnras] {10.1093/mnras/stt1258}, \href
  {https://ui.adsabs.harvard.edu/abs/2013MNRAS.434.3218I} {434, 3218}

\bibitem[\protect\citeauthoryear{{Iglesias-P{\'a}ramo}, {Boselli}, {Gavazzi}
  \& {Zaccardo}}{{Iglesias-P{\'a}ramo} et~al.}{2004}]{Iglesias2004}
{Iglesias-P{\'a}ramo} J.,  {Boselli} A.,  {Gavazzi} G.,   {Zaccardo} A.,  2004,
  \mn@doi [\aap] {10.1051/0004-6361:20034572}, \href
  {https://ui.adsabs.harvard.edu/abs/2004A&A...421..887I} {421, 887}

\bibitem[\protect\citeauthoryear{{Ilbert} et~al.,}{{Ilbert}
  et~al.}{2009}]{Ilbert2009}
{Ilbert} O.,  et~al., 2009, \mn@doi [\apj] {10.1088/0004-637X/690/2/1236},
  \href {https://ui.adsabs.harvard.edu/abs/2009ApJ...690.1236I} {690, 1236}

\bibitem[\protect\citeauthoryear{{Izotov}, {Thuan}  \& {Guseva}}{{Izotov}
  et~al.}{2017}]{Izotov2017}
{Izotov} Y.~I.,  {Thuan} T.~X.,   {Guseva} N.~G.,  2017, \mn@doi [\mnras]
  {10.1093/mnras/stx1629}, \href
  {https://ui.adsabs.harvard.edu/abs/2017MNRAS.471..548I} {471, 548}

\bibitem[\protect\citeauthoryear{{Jaskot} \& {Oey}}{{Jaskot} \&
  {Oey}}{2013}]{Jaskot2013}
{Jaskot} A.~E.,  {Oey} M.~S.,  2013, \mn@doi [\apj]
  {10.1088/0004-637X/766/2/91}, \href
  {https://ui.adsabs.harvard.edu/abs/2013ApJ...766...91J} {766, 91}

\bibitem[\protect\citeauthoryear{{Kennicutt}}{{Kennicutt}}{1992}]{Kennicutt1992}
{Kennicutt} Robert~C. J.,  1992, \mn@doi [\apj] {10.1086/171154}, \href
  {https://ui.adsabs.harvard.edu/abs/1992ApJ...388..310K} {388, 310}

\bibitem[\protect\citeauthoryear{{Kennicutt}}{{Kennicutt}}{1998}]{Kennicutt1998}
{Kennicutt} Robert~C. J.,  1998, \mn@doi [\araa]
  {10.1146/annurev.astro.36.1.189}, \href
  {https://ui.adsabs.harvard.edu/abs/1998ARA&A..36..189K} {36, 189}

\bibitem[\protect\citeauthoryear{{Kewley}, {Geller}, {Jansen}  \&
  {Dopita}}{{Kewley} et~al.}{2002}]{Kewley2002}
{Kewley} L.~J.,  {Geller} M.~J.,  {Jansen} R.~A.,   {Dopita} M.~A.,  2002,
  \mn@doi [\aj] {10.1086/344487}, \href
  {https://ui.adsabs.harvard.edu/abs/2002AJ....124.3135K} {124, 3135}

\bibitem[\protect\citeauthoryear{{Khostovan}, {Sobral}, {Mobasher}, {Best},
  {Smail}, {Stott}, {Hemmati}  \& {Nayyeri}}{{Khostovan}
  et~al.}{2015}]{Khostovan2015}
{Khostovan} A.~A.,  {Sobral} D.,  {Mobasher} B.,  {Best} P.~N.,  {Smail} I.,
  {Stott} J.~P.,  {Hemmati} S.,   {Nayyeri} H.,  2015, \mn@doi [\mnras]
  {10.1093/mnras/stv1474}, \href
  {https://ui.adsabs.harvard.edu/abs/2015MNRAS.452.3948K} {452, 3948}

\bibitem[\protect\citeauthoryear{{Khostovan} et~al.,}{{Khostovan}
  et~al.}{2021}]{khostovan2020}
{Khostovan} A.~A.,  et~al., 2021, \mn@doi [\mnras] {10.1093/mnras/stab778},
  \href {https://ui.adsabs.harvard.edu/abs/2021MNRAS.503.5115K} {503, 5115}

\bibitem[\protect\citeauthoryear{{Koo} \& {Kron}}{{Koo} \&
  {Kron}}{1980}]{Koo1980}
{Koo} D.~C.,  {Kron} R.~T.,  1980, \mn@doi [\pasp] {10.1086/130708}, \href
  {https://ui.adsabs.harvard.edu/abs/1980PASP...92..537K} {92, 537}

\bibitem[\protect\citeauthoryear{{Koyama} et~al.,}{{Koyama}
  et~al.}{2013}]{Koyama2013}
{Koyama} Y.,  et~al., 2013, \mn@doi [\mnras] {10.1093/mnras/stt1035}, \href
  {https://ui.adsabs.harvard.edu/abs/2013MNRAS.434..423K} {434, 423}

\bibitem[\protect\citeauthoryear{{Kriek}, {van Dokkum}, {Labb{\'e}}, {Franx},
  {Illingworth}, {Marchesini}  \& {Quadri}}{{Kriek} et~al.}{2009}]{Kriek2009}
{Kriek} M.,  {van Dokkum} P.~G.,  {Labb{\'e}} I.,  {Franx} M.,  {Illingworth}
  G.~D.,  {Marchesini} D.,   {Quadri} R.~F.,  2009, \mn@doi [\apj]
  {10.1088/0004-637X/700/1/221}, \href
  {https://ui.adsabs.harvard.edu/abs/2009ApJ...700..221K} {700, 221}

\bibitem[\protect\citeauthoryear{{Lange} et~al.,}{{Lange}
  et~al.}{2015}]{Lange2015}
{Lange} R.,  et~al., 2015, \mn@doi [\mnras] {10.1093/mnras/stu2467}, \href
  {https://ui.adsabs.harvard.edu/abs/2015MNRAS.447.2603L} {447, 2603}

\bibitem[\protect\citeauthoryear{{Lee} et~al.,}{{Lee} et~al.}{2009}]{Lee2009}
{Lee} J.~C.,  et~al., 2009, \mn@doi [\apj] {10.1088/0004-637X/706/1/599}, \href
  {https://ui.adsabs.harvard.edu/abs/2009ApJ...706..599L} {706, 599}

\bibitem[\protect\citeauthoryear{{Leitherer} et~al.,}{{Leitherer}
  et~al.}{1999}]{Leitherer1999}
{Leitherer} C.,  et~al., 1999, \mn@doi [\apjs] {10.1086/313233}, \href
  {https://ui.adsabs.harvard.edu/abs/1999ApJS..123....3L} {123, 3}

\bibitem[\protect\citeauthoryear{{Lotz} et~al.,}{{Lotz}
  et~al.}{2017}]{Lotz2017}
{Lotz} J.~M.,  et~al., 2017, \mn@doi [\apj] {10.3847/1538-4357/837/1/97}, \href
  {https://ui.adsabs.harvard.edu/abs/2017ApJ...837...97L} {837, 97}

\bibitem[\protect\citeauthoryear{{Ly} et~al.,}{{Ly} et~al.}{2007}]{Ly2007}
{Ly} C.,  et~al., 2007, \mn@doi [\apj] {10.1086/510828}, \href
  {https://ui.adsabs.harvard.edu/abs/2007ApJ...657..738L} {657, 738}

\bibitem[\protect\citeauthoryear{{Ly}, {Malkan}, {Rigby}  \& {Nagao}}{{Ly}
  et~al.}{2016}]{Ly2016}
{Ly} C.,  {Malkan} M.~A.,  {Rigby} J.~R.,   {Nagao} T.,  2016, \mn@doi [\apj]
  {10.3847/0004-637X/828/2/67}, \href
  {https://ui.adsabs.harvard.edu/abs/2016ApJ...828...67L} {828, 67}

\bibitem[\protect\citeauthoryear{{Madau} \& {Dickinson}}{{Madau} \&
  {Dickinson}}{2014}]{Madau2014}
{Madau} P.,  {Dickinson} M.,  2014, \mn@doi [\araa]
  {10.1146/annurev-astro-081811-125615}, \href
  {https://ui.adsabs.harvard.edu/abs/2014ARA&A..52..415M} {52, 415}

\bibitem[\protect\citeauthoryear{{Maiolino} et~al.,}{{Maiolino}
  et~al.}{2020}]{moons2020}
{Maiolino} R.,  et~al., 2020, \mn@doi [The Messenger]
  {10.18727/0722-6691/5197}, \href
  {https://ui.adsabs.harvard.edu/abs/2020Msngr.180...24M} {180, 24}

\bibitem[\protect\citeauthoryear{{Malhotra} \& {Rhoads}}{{Malhotra} \&
  {Rhoads}}{2004}]{malhotra2004}
{Malhotra} S.,  {Rhoads} J.~E.,  2004, \mn@doi [\apjl] {10.1086/427182}, \href
  {https://ui.adsabs.harvard.edu/abs/2004ApJ...617L...5M} {617, L5}

\bibitem[\protect\citeauthoryear{{Matthee}, {Sobral}, {Santos},
  {R{\"o}ttgering}, {Darvish}  \& {Mobasher}}{{Matthee}
  et~al.}{2015}]{Matthee2015}
{Matthee} J.,  {Sobral} D.,  {Santos} S.,  {R{\"o}ttgering} H.,  {Darvish} B.,
   {Mobasher} B.,  2015, \mn@doi [\mnras] {10.1093/mnras/stv947}, \href
  {http://adsabs.harvard.edu/abs/2015MNRAS.451..400M} {451, 400}

\bibitem[\protect\citeauthoryear{{Matthee}, {Sobral}, {Oteo}, {Best}, {Smail},
  {R{\"o}ttgering}  \& {Paulino-Afonso}}{{Matthee} et~al.}{2016}]{Matthee2016}
{Matthee} J.,  {Sobral} D.,  {Oteo} I.,  {Best} P.,  {Smail} I.,
  {R{\"o}ttgering} H.,   {Paulino-Afonso} A.,  2016, \mn@doi [\mnras]
  {10.1093/mnras/stw322}, \href
  {https://ui.adsabs.harvard.edu/abs/2016MNRAS.458..449M} {458, 449}

\bibitem[\protect\citeauthoryear{{Matthee} et~al.,}{{Matthee}
  et~al.}{2021}]{matthee2021}
{Matthee} J.,  et~al., 2021, arXiv e-prints, \href
  {https://ui.adsabs.harvard.edu/abs/2021arXiv210207779M} {p. arXiv:2102.07779}

\bibitem[\protect\citeauthoryear{{Molnar}, {Ueda}  \& {Umetsu}}{{Molnar}
  et~al.}{2020}]{Molnar2020}
{Molnar} S.~M.,  {Ueda} S.,   {Umetsu} K.,  2020, \mn@doi [\apj]
  {10.3847/1538-4357/abac53}, \href
  {https://ui.adsabs.harvard.edu/abs/2020ApJ...900..151M} {900, 151}

\bibitem[\protect\citeauthoryear{{Moore}, {Lake}  \& {Katz}}{{Moore}
  et~al.}{1998}]{Moore1998}
{Moore} B.,  {Lake} G.,   {Katz} N.,  1998, \mn@doi [\apj] {10.1086/305264},
  \href {https://ui.adsabs.harvard.edu/abs/1998ApJ...495..139M} {495, 139}

\bibitem[\protect\citeauthoryear{{Mundy}, {Conselice}  \& {Ownsworth}}{{Mundy}
  et~al.}{2015}]{Mundy2015}
{Mundy} C.~J.,  {Conselice} C.~J.,   {Ownsworth} J.~R.,  2015, \mn@doi [\mnras]
  {10.1093/mnras/stv860}, \href
  {https://ui.adsabs.harvard.edu/abs/2015MNRAS.450.3696M} {450, 3696}

\bibitem[\protect\citeauthoryear{{Mundy}, {Conselice}, {Duncan}, {Almaini},
  {H{\"a}u{\ss}ler}  \& {Hartley}}{{Mundy} et~al.}{2017}]{Mundy2017}
{Mundy} C.~J.,  {Conselice} C.~J.,  {Duncan} K.~J.,  {Almaini} O.,
  {H{\"a}u{\ss}ler} B.,   {Hartley} W.~G.,  2017, \mn@doi [\mnras]
  {10.1093/mnras/stx1238}, \href
  {https://ui.adsabs.harvard.edu/abs/2017MNRAS.470.3507M} {470, 3507}

\bibitem[\protect\citeauthoryear{{Muzzin}, {van Dokkum}, {Kriek}, {Labb{\'e}},
  {Cury}, {Marchesini}  \& {Franx}}{{Muzzin} et~al.}{2010}]{Muzzin2010}
{Muzzin} A.,  {van Dokkum} P.,  {Kriek} M.,  {Labb{\'e}} I.,  {Cury} I.,
  {Marchesini} D.,   {Franx} M.,  2010, \mn@doi [\apj]
  {10.1088/0004-637X/725/1/742}, \href
  {https://ui.adsabs.harvard.edu/abs/2010ApJ...725..742M} {725, 742}

\bibitem[\protect\citeauthoryear{{Muzzin} et~al.,}{{Muzzin}
  et~al.}{2013}]{Muzzin2013}
{Muzzin} A.,  et~al., 2013, \mn@doi [\apjs] {10.1088/0067-0049/206/1/8}, \href
  {https://ui.adsabs.harvard.edu/abs/2013ApJS..206....8M} {206, 8}

\bibitem[\protect\citeauthoryear{{Naidu}, {Tacchella}, {Mason}, {Bose}, {Oesch}
   \& {Conroy}}{{Naidu} et~al.}{2020}]{naidu2020}
{Naidu} R.~P.,  {Tacchella} S.,  {Mason} C.~A.,  {Bose} S.,  {Oesch} P.~A.,
  {Conroy} C.,  2020, \mn@doi [\apj] {10.3847/1538-4357/ab7cc9}, \href
  {https://ui.adsabs.harvard.edu/abs/2020ApJ...892..109N} {892, 109}

\bibitem[\protect\citeauthoryear{{Oke}}{{Oke}}{1974}]{Oke1974}
{Oke} J.~B.,  1974, \mn@doi [\apjs] {10.1086/190287}, \href
  {http://adsabs.harvard.edu/abs/1974ApJS...27...21O} {27, 21}

\bibitem[\protect\citeauthoryear{{Oteo}, {Sobral}, {Ivison}, {Smail}, {Best},
  {Cepa}  \& {P{\'e}rez-Garc{\'\i}a}}{{Oteo} et~al.}{2015}]{Oteo2015}
{Oteo} I.,  {Sobral} D.,  {Ivison} R.~J.,  {Smail} I.,  {Best} P.~N.,  {Cepa}
  J.,   {P{\'e}rez-Garc{\'\i}a} A.~M.,  2015, \mn@doi [\mnras]
  {10.1093/mnras/stv1284}, \href
  {https://ui.adsabs.harvard.edu/abs/2015MNRAS.452.2018O} {452, 2018}

\bibitem[\protect\citeauthoryear{{Ouchi} et~al.,}{{Ouchi}
  et~al.}{2008}]{Ouchi2008}
{Ouchi} M.,  et~al., 2008, \mn@doi [\apjs] {10.1086/527673}, \href
  {http://adsabs.harvard.edu/abs/2008ApJS..176..301O} {176, 301}

\bibitem[\protect\citeauthoryear{{Ouchi} et~al.,}{{Ouchi}
  et~al.}{2010}]{Ouchi2010}
{Ouchi} M.,  et~al., 2010, \mn@doi [\apj] {10.1088/0004-637X/723/1/869}, \href
  {http://adsabs.harvard.edu/abs/2010ApJ...723..869O} {723, 869}

\bibitem[\protect\citeauthoryear{{Paulino-Afonso}, {Sobral}, {Buitrago}  \&
  {Afonso}}{{Paulino-Afonso} et~al.}{2017}]{PaulinoAfonso2017}
{Paulino-Afonso} A.,  {Sobral} D.,  {Buitrago} F.,   {Afonso} J.,  2017,
  \mn@doi [\mnras] {10.1093/mnras/stw2933}, \href
  {https://ui.adsabs.harvard.edu/abs/2017MNRAS.465.2717P} {465, 2717}

\bibitem[\protect\citeauthoryear{{Penny} \& {Conselice}}{{Penny} \&
  {Conselice}}{2008}]{Penny2008}
{Penny} S.~J.,  {Conselice} C.~J.,  2008, \mn@doi [\mnras]
  {10.1111/j.1365-2966.2007.12535.x}, \href
  {https://ui.adsabs.harvard.edu/abs/2008MNRAS.383..247P} {383, 247}

\bibitem[\protect\citeauthoryear{{P{\'e}rez-Gonz{\'a}lez}
  et~al.,}{{P{\'e}rez-Gonz{\'a}lez} et~al.}{2013}]{Perez2013}
{P{\'e}rez-Gonz{\'a}lez} P.~G.,  et~al., 2013, \mn@doi [\apj]
  {10.1088/0004-637X/762/1/46}, \href
  {http://adsabs.harvard.edu/abs/2013ApJ...762...46P} {762, 46}

\bibitem[\protect\citeauthoryear{{Petrosian}}{{Petrosian}}{1976}]{Petrosian}
{Petrosian} V.,  1976, \mn@doi [\apjl] {10.1086/182301}, \href
  {https://ui.adsabs.harvard.edu/abs/1976ApJ...209L...1P} {210, L53}

\bibitem[\protect\citeauthoryear{{Reddy} et~al.,}{{Reddy}
  et~al.}{2015}]{reddy2015}
{Reddy} N.~A.,  et~al., 2015, \mn@doi [\apj] {10.1088/0004-637X/806/2/259},
  \href {https://ui.adsabs.harvard.edu/abs/2015ApJ...806..259R} {806, 259}

\bibitem[\protect\citeauthoryear{{Rhoads}, {Malhotra}, {Dey}, {Stern},
  {Spinrad}  \& {Jannuzi}}{{Rhoads} et~al.}{2000}]{rhoads2000}
{Rhoads} J.~E.,  {Malhotra} S.,  {Dey} A.,  {Stern} D.,  {Spinrad} H.,
  {Jannuzi} B.~T.,  2000, \mn@doi [\apjl] {10.1086/317874}, \href
  {https://ui.adsabs.harvard.edu/abs/2000ApJ...545L..85R} {545, L85}

\bibitem[\protect\citeauthoryear{{Robertson}, {Ellis}, {Furlanetto}  \&
  {Dunlop}}{{Robertson} et~al.}{2015}]{Robertson2015}
{Robertson} B.~E.,  {Ellis} R.~S.,  {Furlanetto} S.~R.,   {Dunlop} J.~S.,
  2015, \mn@doi [\apjl] {10.1088/2041-8205/802/2/L19}, \href
  {https://ui.adsabs.harvard.edu/abs/2015ApJ...802L..19R} {802, L19}

\bibitem[\protect\citeauthoryear{{Rodighiero} et~al.,}{{Rodighiero}
  et~al.}{2011}]{Rodighiero2011}
{Rodighiero} G.,  et~al., 2011, \mn@doi [\apjl] {10.1088/2041-8205/739/2/L40},
  \href {https://ui.adsabs.harvard.edu/abs/2011ApJ...739L..40R} {739, L40}

\bibitem[\protect\citeauthoryear{{Roediger}, {Bruggen}, {Owers}, {Ebeling}  \&
  {Sun}}{{Roediger} et~al.}{2014}]{Roediger2014}
{Roediger} E.,  {Bruggen} M.,  {Owers} M.~S.,  {Ebeling} H.,   {Sun} M.,  2014,
  \mn@doi [\mnras] {10.1093/mnrasl/slu087}, \href
  {https://ui.adsabs.harvard.edu/abs/2014MNRAS.443L.114R} {443, L114}

\bibitem[\protect\citeauthoryear{{Santos}, {Sobral}  \& {Matthee}}{{Santos}
  et~al.}{2016}]{Santos2016}
{Santos} S.,  {Sobral} D.,   {Matthee} J.,  2016, \mn@doi [\mnras]
  {10.1093/mnras/stw2076}, \href
  {http://adsabs.harvard.edu/abs/2016MNRAS.463.1678S} {463, 1678}

\bibitem[\protect\citeauthoryear{{Santos} et~al.,}{{Santos}
  et~al.}{2020}]{Santos2020}
{Santos} S.,  et~al., 2020, \mn@doi [\mnras] {10.1093/mnras/staa093}, \href
  {https://ui.adsabs.harvard.edu/abs/2020MNRAS.493..141S} {493, 141}

\bibitem[\protect\citeauthoryear{{Sersic}}{{Sersic}}{1968}]{Sersic}
{Sersic} J.~L.,  1968, {Atlas de Galaxias Australes}

\bibitem[\protect\citeauthoryear{{Shen}, {Madau}, {Conroy}, {Governato}  \&
  {Mayer}}{{Shen} et~al.}{2014}]{Shen2014}
{Shen} S.,  {Madau} P.,  {Conroy} C.,  {Governato} F.,   {Mayer} L.,  2014,
  \mn@doi [\apj] {10.1088/0004-637X/792/2/99}, \href
  {https://ui.adsabs.harvard.edu/abs/2014ApJ...792...99S} {792, 99}

\bibitem[\protect\citeauthoryear{{Shipley} et~al.,}{{Shipley}
  et~al.}{2018}]{Shipley2018}
{Shipley} H.~V.,  et~al., 2018, \mn@doi [\apjs] {10.3847/1538-4365/aaacce},
  \href {http://adsabs.harvard.edu/abs/2018ApJS..235...14S} {235, 14}

\bibitem[\protect\citeauthoryear{{Smit}, {Bouwens}, {Labb{\'e}}, {Franx},
  {Wilkins}  \& {Oesch}}{{Smit} et~al.}{2016}]{Smit2016}
{Smit} R.,  {Bouwens} R.~J.,  {Labb{\'e}} I.,  {Franx} M.,  {Wilkins} S.~M.,
  {Oesch} P.~A.,  2016, \mn@doi [\apj] {10.3847/1538-4357/833/2/254}, \href
  {https://ui.adsabs.harvard.edu/abs/2016ApJ...833..254S} {833, 254}

\bibitem[\protect\citeauthoryear{{Sobral} \& {Matthee}}{{Sobral} \&
  {Matthee}}{2019}]{Sobral2019}
{Sobral} D.,  {Matthee} J.,  2019, \mn@doi [\aap]
  {10.1051/0004-6361/201833075}, \href
  {https://ui.adsabs.harvard.edu/abs/2019A&A...623A.157S} {623, A157}

\bibitem[\protect\citeauthoryear{{Sobral} et~al.,}{{Sobral}
  et~al.}{2009}]{Sobral2009}
{Sobral} D.,  et~al., 2009, \mn@doi [\mnras]
  {10.1111/j.1365-2966.2009.15129.x}, \href
  {https://ui.adsabs.harvard.edu/abs/2009MNRAS.398...75S} {398, 75}

\bibitem[\protect\citeauthoryear{{Sobral}, {Best}, {Matsuda}, {Smail}, {Geach}
  \& {Cirasuolo}}{{Sobral} et~al.}{2012}]{Sobral2012}
{Sobral} D.,  {Best} P.~N.,  {Matsuda} Y.,  {Smail} I.,  {Geach} J.~E.,
  {Cirasuolo} M.,  2012, \mn@doi [\mnras] {10.1111/j.1365-2966.2011.19977.x},
  \href {https://ui.adsabs.harvard.edu/abs/2012MNRAS.420.1926S} {420, 1926}

\bibitem[\protect\citeauthoryear{{Sobral}, {Smail}, {Best}, {Geach}, {Matsuda},
  {Stott}, {Cirasuolo}  \& {Kurk}}{{Sobral} et~al.}{2013}]{Sobral2013}
{Sobral} D.,  {Smail} I.,  {Best} P.~N.,  {Geach} J.~E.,  {Matsuda} Y.,
  {Stott} J.~P.,  {Cirasuolo} M.,   {Kurk} J.,  2013, \mn@doi [\mnras]
  {10.1093/mnras/sts096}, \href
  {http://adsabs.harvard.edu/abs/2013MNRAS.428.1128S} {428, 1128}

\bibitem[\protect\citeauthoryear{{Sobral}, {Best}, {Smail}, {Mobasher}, {Stott}
   \& {Nisbet}}{{Sobral} et~al.}{2014}]{Sobral2014}
{Sobral} D.,  {Best} P.~N.,  {Smail} I.,  {Mobasher} B.,  {Stott} J.,
  {Nisbet} D.,  2014, \mn@doi [\mnras] {10.1093/mnras/stt2159}, \href
  {https://ui.adsabs.harvard.edu/abs/2014MNRAS.437.3516S} {437, 3516}

\bibitem[\protect\citeauthoryear{{Sobral}, {Stroe}, {Koyama}, {Darvish},
  {Calhau}, {Afonso}, {Kodama}  \& {Nakata}}{{Sobral}
  et~al.}{2016}]{Sobral2016}
{Sobral} D.,  {Stroe} A.,  {Koyama} Y.,  {Darvish} B.,  {Calhau} J.,  {Afonso}
  A.,  {Kodama} T.,   {Nakata} F.,  2016, \mn@doi [\mnras]
  {10.1093/mnras/stw534}, \href
  {https://ui.adsabs.harvard.edu/abs/2016MNRAS.458.3443S} {458, 3443}

\bibitem[\protect\citeauthoryear{{Sobral} et~al.,}{{Sobral}
  et~al.}{2017}]{Sobral2017}
{Sobral} D.,  et~al., 2017, \mn@doi [\mnras] {10.1093/mnras/stw3090}, \href
  {http://adsabs.harvard.edu/abs/2017MNRAS.466.1242S} {466, 1242}

\bibitem[\protect\citeauthoryear{{Sparre}, {Hayward}, {Feldmann},
  {Faucher-Gigu{\`e}re}, {Muratov}, {Kere{\v{s}}}  \& {Hopkins}}{{Sparre}
  et~al.}{2017}]{Sparre2017}
{Sparre} M.,  {Hayward} C.~C.,  {Feldmann} R.,  {Faucher-Gigu{\`e}re} C.-A.,
  {Muratov} A.~L.,  {Kere{\v{s}}} D.,   {Hopkins} P.~F.,  2017, \mn@doi
  [\mnras] {10.1093/mnras/stw3011}, \href
  {https://ui.adsabs.harvard.edu/abs/2017MNRAS.466...88S} {466, 88}

\bibitem[\protect\citeauthoryear{{Speagle}, {Steinhardt}, {Capak}  \&
  {Silverman}}{{Speagle} et~al.}{2014}]{Speagle2014}
{Speagle} J.~S.,  {Steinhardt} C.~L.,  {Capak} P.~L.,   {Silverman} J.~D.,
  2014, \mn@doi [\apjs] {10.1088/0067-0049/214/2/15}, \href
  {https://ui.adsabs.harvard.edu/abs/2014ApJS..214...15S} {214, 15}

\bibitem[\protect\citeauthoryear{{Stark}, {Ellis}, {Chiu}, {Ouchi}  \&
  {Bunker}}{{Stark} et~al.}{2010}]{Stark2010}
{Stark} D.~P.,  {Ellis} R.~S.,  {Chiu} K.,  {Ouchi} M.,   {Bunker} A.,  2010,
  \mn@doi [\mnras] {10.1111/j.1365-2966.2010.17227.x}, \href
  {https://ui.adsabs.harvard.edu/abs/2010MNRAS.408.1628S} {408, 1628}

\bibitem[\protect\citeauthoryear{{Steidel} \& {Hamilton}}{{Steidel} \&
  {Hamilton}}{1993}]{Steidel1993}
{Steidel} C.~C.,  {Hamilton} D.,  1993, \mn@doi [\aj] {10.1086/116579}, \href
  {https://ui.adsabs.harvard.edu/abs/1993AJ....105.2017S} {105, 2017}

\bibitem[\protect\citeauthoryear{{Steidel}, {Shapley}, {Pettini}, {Adelberger},
  {Erb}, {Reddy}  \& {Hunt}}{{Steidel} et~al.}{2004}]{Steidel2004}
{Steidel} C.~C.,  {Shapley} A.~E.,  {Pettini} M.,  {Adelberger} K.~L.,  {Erb}
  D.~K.,  {Reddy} N.~A.,   {Hunt} M.~P.,  2004, \mn@doi [\apj]
  {10.1086/381960}, \href
  {https://ui.adsabs.harvard.edu/abs/2004ApJ...604..534S} {604, 534}

\bibitem[\protect\citeauthoryear{{Stott}, {Sobral}, {Smail}, {Bower}, {Best}
  \& {Geach}}{{Stott} et~al.}{2013}]{Stott2013}
{Stott} J.~P.,  {Sobral} D.,  {Smail} I.,  {Bower} R.,  {Best} P.~N.,   {Geach}
  J.~E.,  2013, \mn@doi [\mnras] {10.1093/mnras/sts684}, \href
  {https://ui.adsabs.harvard.edu/abs/2013MNRAS.430.1158S} {430, 1158}

\bibitem[\protect\citeauthoryear{{Stroe}, {Sobral}, {Paulino-Afonso}, {Alegre},
  {Calhau}, {Santos}  \& {van Weeren}}{{Stroe} et~al.}{2017}]{Stroe2017}
{Stroe} A.,  {Sobral} D.,  {Paulino-Afonso} A.,  {Alegre} L.,  {Calhau} J.,
  {Santos} S.,   {van Weeren} R.,  2017, \mn@doi [\mnras]
  {10.1093/mnras/stw2939}, \href
  {https://ui.adsabs.harvard.edu/abs/2017MNRAS.465.2916S} {465, 2916}

\bibitem[\protect\citeauthoryear{{Tacchella}, {Dekel}, {Carollo}, {Ceverino},
  {DeGraf}, {Lapiner}, {Mandelker}  \& {Primack Joel}}{{Tacchella}
  et~al.}{2016}]{Tacchella2016}
{Tacchella} S.,  {Dekel} A.,  {Carollo} C.~M.,  {Ceverino} D.,  {DeGraf} C.,
  {Lapiner} S.,  {Mandelker} N.,   {Primack Joel} R.,  2016, \mn@doi [\mnras]
  {10.1093/mnras/stw131}, \href
  {https://ui.adsabs.harvard.edu/abs/2016MNRAS.457.2790T} {457, 2790}

\bibitem[\protect\citeauthoryear{{Villar}, {Gallego}, {P{\'e}rez-Gonz{\'a}lez},
  {Pascual}, {Noeske}, {Koo}, {Barro}  \& {Zamorano}}{{Villar}
  et~al.}{2008}]{Villar2008}
{Villar} V.,  {Gallego} J.,  {P{\'e}rez-Gonz{\'a}lez} P.~G.,  {Pascual} S.,
  {Noeske} K.,  {Koo} D.~C.,  {Barro} G.,   {Zamorano} J.,  2008, \mn@doi
  [\apj] {10.1086/528942}, \href
  {https://ui.adsabs.harvard.edu/abs/2008ApJ...677..169V} {677, 169}

\bibitem[\protect\citeauthoryear{{Weisz} et~al.,}{{Weisz}
  et~al.}{2012}]{Weisz2012}
{Weisz} D.~R.,  et~al., 2012, \mn@doi [\apj] {10.1088/0004-637X/744/1/44},
  \href {https://ui.adsabs.harvard.edu/abs/2012ApJ...744...44W} {744, 44}

\bibitem[\protect\citeauthoryear{{Whitaker} et~al.,}{{Whitaker}
  et~al.}{2014}]{Whitaker2014}
{Whitaker} K.~E.,  et~al., 2014, \mn@doi [\apj] {10.1088/0004-637X/795/2/104},
  \href {https://ui.adsabs.harvard.edu/abs/2014ApJ...795..104W} {795, 104}

\bibitem[\protect\citeauthoryear{{Yang} et~al.,}{{Yang}
  et~al.}{2017}]{Yang2017}
{Yang} H.,  et~al., 2017, \mn@doi [\apj] {10.3847/1538-4357/aa7d4d}, \href
  {https://ui.adsabs.harvard.edu/abs/2017ApJ...844..171Y} {844, 171}

\bibitem[\protect\citeauthoryear{{Zhou} et~al.,}{{Zhou}
  et~al.}{2019}]{Zhou2019}
{Zhou} Y.,  et~al., 2019, \mn@doi [\apj] {10.3847/1538-4357/ab1b4b}, \href
  {https://ui.adsabs.harvard.edu/abs/2019ApJ...877..104Z} {877, 104}

\bibitem[\protect\citeauthoryear{{de Jong} et~al.,}{{de Jong}
  et~al.}{2019}]{4most2019}
{de Jong} R.~S.,  et~al., 2019, \mn@doi [The Messenger]
  {10.18727/0722-6691/5117}, \href
  {https://ui.adsabs.harvard.edu/abs/2019Msngr.175....3D} {175, 3}

\makeatother
\end{thebibliography}







\bsp	
\label{lastpage}
\end{document}